\documentclass[%
 amsmath,amssymb,
 aps, physrev,
]{revtex4-2}

\usepackage{graphicx}
\usepackage{dcolumn}
\usepackage{bm}
\usepackage[colorlinks = true,
            linkcolor = red,
            urlcolor  = blue,
            citecolor = magenta,
            anchorcolor = green]{hyperref}

\usepackage{graphicx, caption, subcaption}

\usepackage{acro}

\DeclareAcronym{qcd}{
  short=QCD,
  long=quantum chromodynamics,
}

\DeclareAcronym{njl}{
  short=NJL,
  long=Nambu--Jona-Lasinio,
}
\DeclareAcronym{rmf}{
  short=RMF,
  long=relativistic mean field,
}
\DeclareAcronym{eos}{
  short=EOS,
  long=equation of state,
}
\DeclareAcronym{ns}{
  short=NS,
  long=Neutron star,
}
\DeclareAcronym{hs}{
  short=HS,
  long=hybrid \ac{ns},
}
\DeclareAcronym{nro}{
  short=NRO,
  long=non-radial oscillation,
}
\DeclareAcronym{tov}{
  short=TOV,
  long=Tolmann-Oppenheimer-Volkoff,
}
\DeclareAcronym{nsm}{
  short=NSM,
  long=neutron star matter,
}
\DeclareAcronym{hqpt}{
  short=HQPT,
  long=hadron-quark phase transition,
}
\DeclareAcronym{gw}{
  short=GW,
  long=gravitational wave,
}

\DeclareAcronym{cp}{
  short=CP,
  long=charge-conjugation and parity,
}
\DeclareAcronym{pq}{
  short=PQ,
  long=Peccei-Quinn,
}

\DeclareAcronym{hic}{
  short=HIC,
  long=heavy ion collision,
}

\usepackage{xcolor}

\usepackage{IEEEtrantools}

\begin{document}

\def\be{\begin{equation}}
\def\ee{\end{equation}}
\def\bearr{\begin{eqnarray}}
\def\eearr{\end{eqnarray}}
\def\zbf#1{{\bf {#1}}}
\def\bfm#1{\mbox{\boldmath $#1$}}
\def\hf{\frac{1}{2}}
\def\sl{\hspace{-0.15cm}/}
\def\omit#1{_{\!\rlap{$\scriptscriptstyle \backslash$}
{\scriptscriptstyle #1}}}
\def\vec#1{\mathchoice
        {\mbox{\boldmath $#1$}}
        {\mbox{\boldmath $#1$}}
        {\mbox{\boldmath $\scriptstyle #1$}}
        {\mbox{\boldmath $\scriptscriptstyle #1$}}
}

\title{CP violation in cold dense quark matter and axion effects on the non-radial oscillations of neutron stars}

\author{Deepak Kumar$^{1}$}
\email{deepak.kumar@iopb.res.in}
\author{Hiranmaya Mishra$^{2}$}
\email{hiranmaya@niser.ac.in}

\affiliation{$^{1}$Institute of Physics, Sachivalaya Marg, Bhubaneswar 751005,India 
}
\affiliation{$^{2}$School of Physical Sciences,National Institute of Science Education and Research, An OCC of Homi Bhabha National Institute, Jatni 752050, India}

\date{\today} 

\begin{abstract}
Charge-conjugation and parity violation in strong interaction for cold dense quark matter is studied with axions of quantum chromodynamic within the three flavor Nambu--Jona-Lasinio model that includes the coupling of axions to quarks. We first calculate the effective potential for axions at finite baryon density and zero temperature including the effects of a first-order chiral phase transition. Using the equation of state for quark matter with axions and a hadronic matter equation of state in the ambit of a relativistic mean field theory in quantum hadrodynamics, we discuss the hadron-quark phase transition. We use a Gibbs construct for the same satisfying the constraints of beta equilibrium and charge neutrality as appropriate for the neutron star matter. The equation of state so obtained is used to investigate the structure of hybrid neutron stars. It is found that with the presence of axions, it is possible to have stable hybrid neutron stars having an inner core of quark matter both in pure quark matter phase as well as in a mixed phase with hyperonic matter along with an outer core of hyperonic matter and is in agreement with modern astrophysical constraints. We also discuss the properties of non-radial oscillations of such hybrid neutron stars. It is observed that the quadrupolar fundamental modes ($f$-modes) for such hybrid neutron stars get substantial enhancements both due to a larger quark core in the presence of axions and from the hyperons as compared to a canonical nucleonic neutron stars.
\end{abstract}

\maketitle



\section{\label{sec:introduction} Introduction}\label{intro}
\ac{ns}s are exciting cosmic laboratories to explore the dense region of \ac{qcd} phase diagram that cannot be explored in the current experiments with heavy ion collisions or theoretically using lattice \ac{qcd} simulations. Gross structural properties of such \ac{ns}s like mass, radius, moment of inertia, tidal deformability of binary merging systems as well as different oscillation modes depend crucially on the composition of constituent matter which affect \ac{eos} of the matter inside \ac{ns}. Indeed, the \ac{gw}s from binary \ac{ns} merger and associated measurement of tidal deformability constrains \ac{nsm} \ac{eos} \cite{LIGOScientific:2018cki, LIGOScientific:2018hze}. Along with it, the high precision x-ray mission like Neutron star Interior Composition Explorer (NICER) has  played a key role in providing a tighter constraint on the mass-radius relation thereby also on \ac{eos} of strongly interacting matter \cite{Miller:2019cac, Riley:2019yda, Riley:2021pdl}.

While strong interaction is known to respect space-time reflection symmetry to a very high degree, this is not a direct consequence of laws of \ac{qcd}, which, in principle permits a \ac{cp} violating topological term 
\begin{equation}
{\cal L}_\theta=\frac{\theta}{32\pi^2} g^2 F_{\mu\nu}^a\tilde F^{a\mu\nu}.
\label{lth}
\end{equation}
In the above, $g$ is the strong coupling,  $F_{\mu\nu}^a$ is the gluon field strength with $\tilde F^{a\mu\nu}$ being its dual and $\theta$ is the \ac{qcd} vacuum angle. Such a term, while being consistent with Lorentz invariance and gauge invariance, violates \ac{cp} unless $\theta=0$ mod $\pi$. For \ac{qcd} vacuum, spontaneous parity violation does not arise for $\theta=0$ \cite{Vafa:1984xg}. On the other hand for $\theta=\pi$ there could be spontaneous \ac{cp} violation by the Dashen phenomenon \cite{Dashen:1970et} as the Lagrangian is explicitly \ac{cp} conserving but spontaneous \ac{cp} violation can arise as there are two degenerate vacua which differ by a \ac{cp} transformation from each other. The \ac{cp} symmetry conserving nature of \ac{qcd} has been established by precise experiments that sets limit on the intrinsic electric dipole moment of neutron. The current experimental limit on this leads to a limit on the coefficient of the \ac{cp} violating term of the \ac{qcd} Lagrangian as $\theta < 0.7 \times 10^{-11}$ \cite{Baker:2006ts}. This apart, various experimental studies on pseudoscalar mass ratios \cite{Kawarabayashi:1980uh, Pendlebury:2015lrz} as well as lattice \ac{qcd} simulations \cite{Guo:2015tla, Bhattacharya:2015esa} also conclude that the value of $\theta$ is indeed very close to zero. This smallness of \ac{cp} violation term or its complete absence is not understood completely though a possible explanation is given in terms of spontaneous breaking of a new symmetry : the \ac{pq} symmetry \cite{Peccei:1977ur}. While the \ac{pq} mechanism is an elegant and robust method to solve the strong \ac{cp} problem in a dynamical manner which predicts the smallness of the $\theta$, spontaneous breaking of the \ac{pq} symmetry \cite{Peccei:1977hh,Peccei:1977ur} also naturally predicts a pseudo-Goldstone boson field (`$a$') which is known as axion \cite{GrillidiCortona:2015jxo, Kim:2008hd}. This represents the quantum fluctuation of axion field around its vacuum expectation value $\langle a\rangle$ with the identification $\theta = \langle a \rangle / f_a$, where $f_a$ is the axion decay constant. It may be noted that even if \ac{cp} is not violated for \ac{qcd} vacuum, it is possible that it can be violated for \ac{qcd} matter at finite temperature or density.

Axions are weakly interacting and are very light and therefore are also good bosonic dark matter candidates \cite{Weinberg:1977ma, Turner:1990uz, Duffy:2009ig, Visinelli:2009zm}. In the context of \ac{ns}s, axions can potentially be gravitationally trapped in the core of \ac{ns}s during their stellar formation as well as possible continual accumulation and can influence several structural properties of \ac{ns}s like mass, tidal deformability \cite{Karkevandi:2021ygv}. One of the main scope of the current investigation is to study the effects of axions on \ac{ns} macroscopic properties and its oscillations.

It may be noted that the use of perturbative methods in studying axion effects on dense matter for the baryon densities relevant for \ac{ns}s is questionable. Therefore in the low energy regime, where, nonperturbative effects are important, it is necessary to take resort to \ac{qcd} like models and effective field theories to explore the dense matter physics. A frequently used effective theory is the chiral perturbation theory ($\chi$PT) which plays an important role to study vacuum structure of \ac{qcd} as well as axion properties at low energies by systematically expanding in powers of momenta of light mesons \cite{Bernard:2012ci, Metlitski:2005di}. Such a scheme is very useful and advantageous at low energies. Indeed, the topological susceptibility as vanishing temperature that is predicted here matches with lattice \ac{qcd} results rather remarkably \cite{Bonati:2015vqz}. However, at large temperatures and/or densities its applicability diminishes as it lacks information regarding \ac{qcd} phase transition. This necessitates the use of \ac{qcd} like models that incorporate axions and capture the \ac{qcd} phase transition dynamics. This has been attempted using linear sigma model \cite{Mizher:2008dj} and \ac{njl} model and its various extensions \cite{Frank:2003ve, Boomsma:2009eh, Boomsma:2009yk, Chatterjee:2011yz, Chatterjee:2014csa, Fukushima:2001hr, Sakai:2011gs}.

In the present investigation, we wish to study the effects of axions in \ac{hqpt} in the context of \ac{ns}s. We adopt the three-flavor local \ac{njl} model as an effective model for chiral symmetry breaking in strong interactions. Specifically, we shall use a formulation of the model that includes an instanton induced interaction which breaks U(1)$_A$ symmetry and describe both the spontaneous breaking of chiral symmetry and interaction of axions with quarks. The \ac{njl} model has been used earlier to study the \ac{cp} violating effects and effect of $\theta$ vacuum on the \ac{qcd} phase diagram \cite{Frank:2003ve, Boomsma:2009eh, Boomsma:2009yk, Chatterjee:2011yz, Chatterjee:2014csa}. This approach has been considered in Refs. \cite{Lu:2018ukl} and \cite{Bandyopadhyay:2019pml, Ali:2020jsy} to study the axion mass and self-coupling at finite temperature in absence as well as in the presence of a magnetic field respectively for 2-flavor as well as for  three flavors Polyakov loop extended \ac{njl} (P\ac{njl}) model \cite{Das:2020pjg}. In this context, compact stars bearing axions within a 3-flavor scenario has been recently studied in Ref. \cite{Lopes:2022efy} where it was shown that the presence of axions stabilizes massive hybrid neutron stars (HS) against gravitational collapse by weakening the \ac{hqpt} with a smaller critical density.

It may be noted here that while tidal deformability extracted from the phases of \ac{gw} front in GW170817 event puts constraints on \ac{eos} of merging \ac{ns}s \cite{Radice:2020ddv, Raithel:2019uzi}. Within the current observational status it is difficult to distinguish between a canonical \ac{ns} i.e. without a quark matter core or a \ac{hs} with a quark matter core or a core of quark matter in a mixed phase with hadronic matter and is  known as the 'masqueraded' problem \cite{Alford:2004pf}. In this context, it is suggested that the study of \ac{nro} modes of \ac{ns} can have a possibility of getting the compositional information of matter inside a \ac{ns}. This is essentially due to the fact the \ac{nro}s not only depend upon \ac{eos} i.e. pressure as a function of the energy density ($p(\epsilon)$) but also its derivative ($dp/d\epsilon$) \cite{Kumar:2021hzo, McDermott:1983}. The different \ac{nro}s of \ac{ns}s are characterized by the restoring forces that bring back the perturbed star to equilibrium \cite{Pradhan:2022vdf}. The important modes of these oscillations are the pressure ($p$) modes, fundamental ($f$) modes and gravity ($g$) modes. The frequency of $g$ modes is lower than that of $p$ modes while the same for $f$ modes lies in between. Several of these modes are expected to be excited during supernovae \cite{Radice:2018usf}, or in during the star quake \cite{Rencoret:2021zmn, Keer:2014uva}, or in isolated perturbed \ac{ns} \cite{Doneva:2013zqa} or during post merger phase of binary \ac{ns}s \cite{Stergioulas:2011gd, Bauswein:2011tp, Takami:2014zpa}. Both the $f$ and $g$ modes are correlated with the tidal deformability but the $g$ modes are too weak to be detected by present \ac{gw} detectors. The focus of our current work is on the characteristic changes in $f$ mode oscillations due to the changes in the microscopic quark matter \ac{eos} when strong \ac{cp} violating effects are included through the axion fields. These modes are predicted to produce significant amount of gravitational energy when the neutron star is unstable. Further, during the inspiral stage of neutron star mergers spin and eccentricity enhance the excitations of f-modes \cite{Chirenti:2016xys, Steinhoff:2021dsn}. These modes also lie within the sensitivity of current and upcoming gravitational wave detectors \cite{Pratten:2019sed}.

We organize the paper in the following manner. In section \ref{strong_cp_violation} we discuss the three flavors \ac{eos} model with a \ac{cp} violating term and write down the resulting \ac{eos} in terms of condensates in scalar and pseudoscalar channels. While the former condensate arise without the \ac{cp} violating terms, the latter arises when the \ac{cp} violating parameter is non-vanishing. In section \ref{eos_hm_rmf_model} we discuss a \ac{rmf} for hadronic matter including hyperonic degrees of freedom. In section \ref{hqpt} we discuss the \ac{hqpt} at densities which are possible at the core of  large mass \ac{ns}s.In section \ref{neutron_star_and_non_radial_oscillation} we write down the equations for the fluid perturbation from which we derive the oscillation frequencies. In section \ref{results_and_discussion} we discuss the results in some detail. In section \ref{summary_and_conclusion}, we summarize the salient features of the present investigation and discuss the outlook.

\section{Strong CP violation at finite density within NJL model} \label{strong_cp_violation}
We shall consider the spontaneous \ac{cp} violation in the dense matter within the ambit of \ac{njl} model. Explicitly, the Lagrangian density of the three flavour P\ac{njl} model with the Kobayashi-Maskawa-’t Hooft determinant interaction term incorporating the interaction with the axion can be expressed as \cite{Sakai:2011gs, Boomsma:2009eh, Boomsma:2009eh, Chatterjee:2014csa, Boomsma:2009yk}
\begin{IEEEeqnarray}{rCl}
\mathcal{L} &=&  \bar{q}(i \gamma^\mu {\partial_\mu} - \hat{m})q + G_s \sum_{A=0}^8\big[ (\bar{q}\lambda^A q)^2 + (\bar{q}i\gamma_5 \lambda^A q)^2 \big] \nonumber \\
&& - K \big[ e^{i\theta} \text{det}\lbrace \bar{q}(1+\gamma^5) q \rbrace + e^{-i\theta} \text{det} \lbrace\bar{q}(1-\gamma^5) q\rbrace\big] - G_v\left[(\bar q\gamma^\mu q)^2+(\bar q\gamma^\mu\gamma^5 q)^2\right].
\label{lagrangian}
\end{IEEEeqnarray}
Here $q=(q_u,q_d,q_s)^T$ is the quark field, $\hat{m}$ represents the current quark mass matrix $\text{diag}(m_{u},m_{d},m_{s})$. In the present investigation we consider $m_{u}=m_{d}=m_0$. $\lambda^0 = \sqrt{2/3} ~I_{3\times3}$, here $I_{3\times3}$ is the $3\times 3$ identity matrix in flavor space, $\lambda^A$ with 
$A=1,2,...,8$ are the Gell Mann matrices in flavor space. The parameter $G_s$ denotes the coupling of the four-quark interaction which includes scalar and pseudoscalar type interactions. This interaction term is symmetric under $SU(3)_{L}\times SU(3)_{R}\times U(1)_V\times U(1)_A\times SU(3)_C$ 
symmetry. $K$ is the coupling of the Kobayashi-Maskawa -'t Hooft determinant interaction. 
This determinant is taken in the flavor space. The determinant interaction term explicitly breaks the $U(1)_A$ symmetry of the Lagrangian. This interaction is important for obtaining for the mass splitting between the pseudo-scalar isosinglet mesons $\eta^{\prime}$ and $\eta$. Inclusion of this interaction term makes it possible to reproduce the mass values of $\eta$ and $\eta^{\prime}$ within the framework of the \ac{njl} model. The effects of axions are incorporated in this determinant interaction among the quark through the term $\theta \equiv \langle a\rangle/f_a$. We shall consider here the case where the U(1) \ac{pq} symmetry breaking occurs at the grand unified scale so that $f_a \sim 10^{15}\ {\rm GeV}$ giving rise to a very light and weakly interacting axion \cite{Kim:1979if, Dine:1981rt}. Thus we can take the axion field $a$ as its vacuum expectation value. The last term in the Lagrangian in Eq. (\ref{lagrangian}) represents the vector and axial-vector channels of interaction with a positive coupling constant $G_v$. Let us note that the vector coupling term is taken as a flavor singlet type coupling. One could take a vector octet type of coupling by including vector terms of the type $G_v \bar{q}\gamma^\mu\lambda_a q$.

The thermodynamic potential with the above Lagrangian can be calculated in terms of quark antiquark condensates in the
scalar and pseudo-scalar channels. This was done using an explicit variational construct 
for the ground state in Ref. \cite{Chatterjee:2014csa} and including the effect of magnetic fields Ref. \cite{Chatterjee:2011yz} without the vector coupling. Following a similar procedure, the thermodynamic potential (negative of pressure) can be written for $G_v = 0$ as, \cite{Chatterjee:2014csa, Lopes:2022efy} 

\begin{IEEEeqnarray}{rCl}
\Omega_{}(I_s^i, I_p^i,\theta,T,\mu) &=&  \Omega_{\bar{q}q} +  \sum_{i=u,d,s} 2 G_s (I_s^i{}^2 + I_p^i{}^2) + 4 K (\cos \theta I_s^uI_s^dI_s^s+\sin \theta I_p^uI_p^dI_p^s) \nonumber \\
&& - 4 K \bigg(\cos \theta (I_s^uI_p^dI_p^s+ I_s^dI_p^uI_p^s+ I_s^sI_p^dI_p^u) + \sin \theta (I_p^uI_s^dI_s^s+I_p^dI_s^uI_s^s+I_p^sI_s^uI_s^d)\bigg) \label{thpot}      
\end{IEEEeqnarray}

\noindent Here $I_s^i =- \langle\bar{q}^i q^i \rangle$ is the scalar condensate for the flavor $i(i=u,d,s)$ and $I_p^i= \langle\bar{q}^ii\gamma_5 q^i\rangle$ is the pseudo-scalar condensate for flavor $i(i=u,d,s)$. Further, $\Omega_{\bar q q}$ is the contribution
\begin{IEEEeqnarray}{rCl}
\Omega_{\bar q q}&=&-\frac{2N_c}{(2\pi)^3}\sum_{i=u,d,s}\int d\zbf p E^i(\zbf p) - \frac{2N_c}{(2\pi)^3}\sum_{i=u,d,s}\int d\zbf p \left[\log(1+e^{-\beta((E^i(\zbf p)-\mu^i))}+\log(1+e^{-\beta(E^i(\zbf p)+\mu^i)})\right],
\label{omgqbq}
\end{IEEEeqnarray}
with $\beta$=1/T. The dispersion relation for the quarks are given for i-th flavor as
$E^i(\zbf p)=\sqrt{\zbf p^2+M^i{}^2}$ with the mass of the $i$-th flavor quark M$^i$=$\sqrt{M_s^i{}^2+M_p^{i}{}^2}$ where $M_s^i$ and $M_p^i$ are, the contributions to the constituent quark mass (for ith flavor) from the scalar and pseudoscalar condensates respectively and they are given  for each flavor as
\begin{IEEEeqnarray}{rCl}
M_s^u &=& m_u + 4G_s I_s^u + 2K\left(\cos\theta(I_s^dI_s^s - I_p^dI_p^s) - \sin\theta(I_p^dI_s^s + I_b^sI_s^d)\right), \label{msu} \\
M_s^d &=& m_d + 4G_s I_s^d + 2K\left(\cos\theta(I_s^uI_s^s - I_p^uI_p^s) - \sin\theta(I_p^uI_s^s + I_p^sI_s^u)\right), \label{msd} \\
M_s^s &=& m_s + 4G_s I_s^s + 2K\left(\cos\theta(I_s^uI_s^d - I_p^uI_p^d) - \sin\theta(I_p^uI_s^d + I_p^dI_s^u)\right), \label{mss}
\end{IEEEeqnarray}
for the scalar components while  the pseudoscalar components are given by
\begin{IEEEeqnarray}{rCl}
M_p^u &=& 4G_s I_p^u - 2K\left(\cos\theta(I_s^dI_p^s + I_p^dI_s^s) + \sin\theta(I_s^dI_s^s - I_p^dI_p^s)\right), \label{mpu} \\
M_p^d &=& 4G_s I_p^d - 2K\left(\cos\theta(I_s^sI_p^u + I_p^sI_s^u) + \sin\theta(I_s^sI_s^u - I_p^sI_p^u)\right), \label{mpd} \\
M_p^s &=& 4G_s I_p^s - 2K\left(\cos\theta(I_s^uI_p^d + I_p^uI_s^d) + \sin\theta(I_s^uI_s^d - I_p^uI_p^d)\right). \label{mps}
\end{IEEEeqnarray}
In the above $I_s^i$ and $I_p^i$ ,($i=u,d,s$) are the condensates in the scalar and pseudo-scalar channels respectively for the $i$-th flavor and are given explicitly as
\begin{IEEEeqnarray}{rCl}
I_s^i\equiv -\langle\bar q_i q_i\rangle &=& \frac{2N_c}{(2\pi)^3}\int d\zbf p \frac{M_s^i}{E^i(\zbf p)} \left(1-\sin^2\theta^i_-(\zbf p,\beta,\mu^i)-\sin^2\theta^i_+(\zbf p,\beta,\mu^i)\right),
\label{isi}
\\
I_p^i\equiv \langle\bar q_ii\gamma_5 q_i\rangle &=& \frac{2N_c}{(2\pi)^3}\int d\zbf p \frac{M_p^i}{E^i(\zbf p)} \left(1-\sin^2\theta^i_-(\zbf p,\beta,\mu^i)-\sin^2\theta^i_+(\zbf p,\beta,\mu^i)\right).
\label{ipi}
\end{IEEEeqnarray}
Here, the functions $\sin^2\theta_\mp$ are the thermal distribution functions given as
\begin{IEEEeqnarray}{rCl}
\sin^2\theta^i_\mp &=& \frac{1}{\exp\left(\beta(E^i(\zbf p)\mp\mu^i)\right)}.
\label{distbn}
\end{IEEEeqnarray}

Thus the thermodynamic potential or the 'effective potential' as a function of axion parameter $\langle a\rangle/f_a\equiv \theta$, and condensates $I_s$ and $I_p$ as given temperature and chemical potential as given in Eq. (\ref{thpot}) gets completely defined. The behavior of the effective potential for axions for finite temperatures and vanishing chemical potential have been analyzed earlier within the ambit of \ac{njl} model \cite{Chatterjee:2011yz}. We shall focus our attention here however for vanishing temperature and finite chemical potential
which is relevant for compact stars.

At vanishing temperature the the distribution functions for antiparticle $\sin^2\theta_+^i$ vanishes while that for particles reduces to Heaviside ($\Theta$) functions i.e. $\sin^2\theta_-^i(\zbf p,\mu^i,T=0)=\Theta(\mu^i-E^i(\zbf p))$. Further, using the identity $\lim_{a\rightarrow \infty\ln(1+\exp(-ax))}/a=-x\Theta(-x)$, in Eq. (\ref{omgqbq}), the zero temperature contribution $\Omega_{\bar q q}$ become
\be
\Omega_{\bar q q}=-\frac{N_c}{\pi^2}\sum_{i}\left[\Lambda^4H\left(\frac{M^i}{\Lambda}\right)-k_f^i{}^4H\left(\frac{M^i}{k_f^i}\right)\right]-\sum_i\mu^in^i.
\label{omgqbarqt0}
\ee
where, $\Lambda$ is the three momentum cut off used in the \ac{njl} model and we 
we have introduced the dimensionless function $H(z)$ as
\be
H(z)=\frac{1}{2}(1+z^2)^{3/2}-\frac{z^2}{8}(1+z^2)^{1/2}-\frac{z^4}{8}\ln\left(\frac{1+\sqrt{1+z^2}}{z}\right).
\label{hz}
\ee
Further, in Eq. (\ref{omgqbarqt0}) the fermi-momentum $k_f^i$ of each flavor  given as  $k_f^i=\sqrt{\mu^i{}^2-M^i{}^2}$ and the density of each quark species  as $n^i=\frac{k_f^i{}^3}{\pi^2}$. As defined earlier the mass $M^i=\sqrt{M_s^i{}^2+M_p^i{}^2}$ with the contributions 
from the quark anti-quark condensates in the scalar channel to the mass $M_s^i$ as given in Eqs. (\ref{msu} - \ref{mss}) and the pseudo-scalar channel to the mass $M_p^i$ as given in Eqs. (\ref{mpu} - \ref{mps}). At zero temperature, the condensates $I_s$ and $I_p$ given in Eqs. (\ref{isi}) and Eq. (\ref{ipi}) reduce to
\begin{IEEEeqnarray}{rCl}
I_s^i=M_s^i\frac{N_c}{\pi^2}\left[\Lambda^2 G\left(\frac{M^i}{\Lambda}\right)-k_f^i{}^2 G\left(\frac{M^i}{k_f^i}\right)\right],
\label{isit0}
\\
I_p^i=M_p^i\frac{N_c}{\pi^2}\left[\Lambda^2 G\left(\frac{M^i}{\Lambda}\right)-k_f^i{}^2 G\left(\frac{M^i}{k_f^i}\right)\right],
\label{ipit0}
\end{IEEEeqnarray}
with,
\begin{IEEEeqnarray}{rCl}
G(z)=\frac{1}{2}\left[(1+z)^2-z^2\ln\left(\frac{1+\sqrt{1+z^2}}{z}\right)\right].
\label{gz}
\end{IEEEeqnarray}

The difference of the vacuum energy densities between the non-perturbative vacuum (with corresponding constituent quark mass $M^i$) 
and the energy density of the perturbative vacuum (with the corresponding current quark mass) for $\theta=0$ is the bag constant $B_0$ i.e.
\be
B_0=\Omega(I_s^i,I_p^i,\theta=0,T=0,\mu^i=0)-\Omega(I_s^i{}^0,I_p^i=0,T=0,\mu^i=0),
\ee
which needs to be subtracted from Eq.(\ref{thpot}) so that the thermodynamic potential vanishes at vanishing temperature and density for $\theta=0$.
On the other hand, one can define a $\theta$-dependent bag constant $B_\theta$ as
\be
B_\theta=\Omega(I_s^i,I_p^i,\theta,T=0,\mu^i=0)-\Omega(I_s^i{}^0,I_p^i=0,\theta, T=0,\mu^i=0),
\ee
where, $I_s^i{}^0$ correspond to the scalar condensate value for the i$th$ flavor with current quark mass $m_i$ at zero temperature and zero baryon density. The pressure in the quark phase, $p_{\rm QP}$ which is negative of thermodynamic potential, is given as

\be
p_{\rm QP}(\theta,I^i_s,I_p^i, \mu^i)=-\Omega(I^i_s,I^i_p,\theta,T=0,\mu^i)+B_\theta. \label{pressure_qm}
\ee
From standard thermodynamic relation, the energy density of quarks is given as
\be
\epsilon_{\rm QP}(\theta,I^i_s,I_p^i, \mu^i)=\sum_i\mu_in_i-p_{\rm QP}(\theta,I^i_s,I_p^i, \mu^i).
\ee

Further \ac{ns} matter needs to be electrically charge neutral as well as $\beta$ equilibrated. With  the later condition, the chemical potentials of the quarks can be expressed in terms
of baryon chemical potential $\mu_{\rm B}$ and electric charge chemical potential $\mu_{\rm E}$ as $\mu_i=\mu_{\rm B}/3+q_i\mu_{\rm E}$ where, $q_i$ 
is the electric charge of the $i^{\rm th}$ species of quark. In the present notation the electron chemical potential is $\mu_e=-\mu_{\rm E}$.
The condition of charge neutrality is given by
\be
\frac{2}{3}n_u-\frac{1}{3}(n_d+n_s)-n_e=0 \label{charge_neutrality_qm},
\ee
where, the number densities of each quark species is defined after Eq.(\ref{hz}) and the 
electron number density is given as $n_e=k_f^e{}^3/(3\pi^2)$, with, $k_f^e=\sqrt{\mu_e^2-m_e^2}$,
$m_e$ being the electron mass.

For numerical evaluation of the pressure and other thermodynamic quantities, the parameters of the three flavors \ac{njl} model are chosen as follows. Let us note that the coupling constant $G_s$ has dimension [M]$^{-2}$ while the determinant coupling $K$ has a dimension [M]$^{-5}$. Further to regularize the divergent integral we use a sharp cut off $\Lambda$ in the three momentum space. Thus, we have five parameters in total - namely, the three current quark masses and the two couplings $G_s$ and $K$.We have chosen the parameters as $\Lambda$=602.3 MeV; $G_s\Lambda^2$=1.835; $K\Lambda^5=12.36$; $m_u=m_d$=5.5 MeV and $m_s$=140.7 as was used in Ref. \cite{Rehberg:1995kh}. After choosing the $m_u=m_d=5.5$MeV, the remaining four parameters were  fixed by fitting the pion decay constant and the masses of pion, kaon and $\eta^\prime$. With these parameters the mass of $\eta$ is underestimated by 6$\%$ and the masses of the light quarks turn out to be $M^{u,d}$=368 MeV while that of strange quark turns out to be $M^s=549$ MeV at zero temperature and vanishing density with $\theta=0$. 

As noted earlier, the above calculations have been done for vanishing value of vector coupling $G_v$. The effect of finite $G_v$ lies in the modification of the chemical potential the quarks as
\begin{IEEEeqnarray}{rCl}
\tilde{\mu}_i = \mu_i - 2 G_v \sum_{i=u,d,s} n_i.
\end{IEEEeqnarray}
\noindent Further, a finite $G_v$ also lead to an extra term in the thermodynamic potential given in Eq. (\ref{thpot}) i.e.
\begin{IEEEeqnarray}{rCl}
\Omega_{\rm Tot} = \Omega(I_s^i,I_p^i) - G_v n^2,
\end{IEEEeqnarray}
where $n = \sum_{i = u,d,s} n_i$ is the total quark matter density. 

\begin{figure}
     \begin{subfigure}[t]{0.49\textwidth}
      \includegraphics[width=\textwidth]{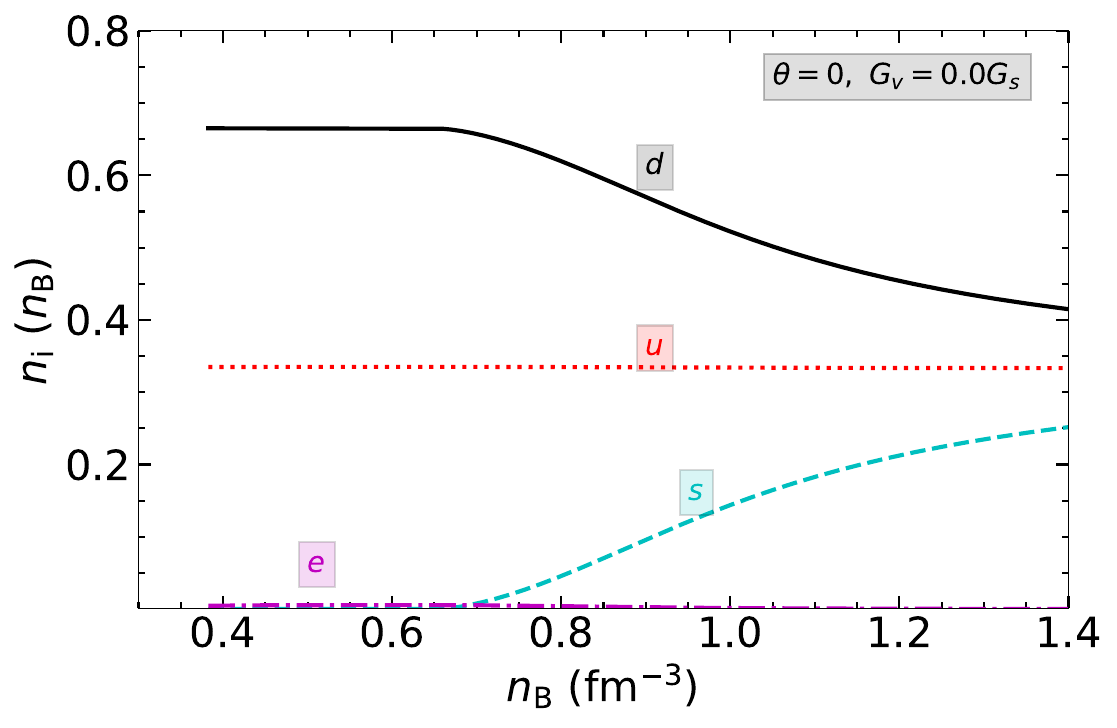}
      \caption{}
    \end{subfigure}
    \begin{subfigure}[t]{0.49\textwidth}
      \includegraphics[width=\textwidth]{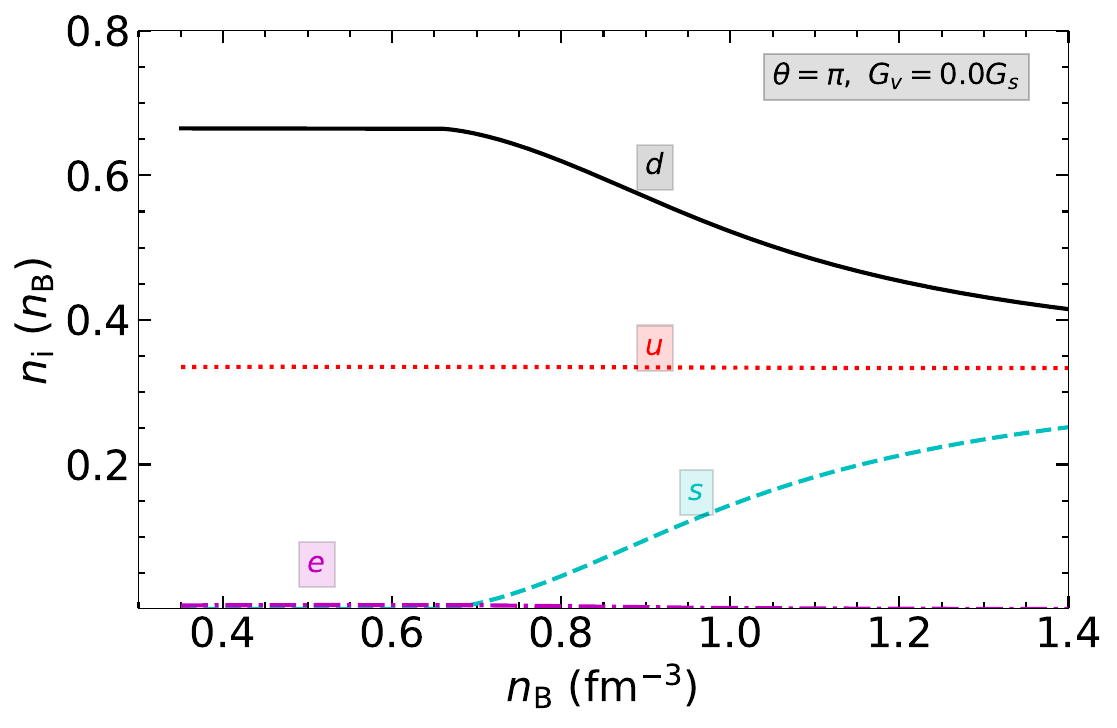}
      \caption{}
    \end{subfigure}
\caption{\label{fig:uds_fraction} Particle densities in charge neutral three flavors quark matter as a function of baryon density for (a) without axion ($\theta = 0$)  and (b) with axion ($\theta = \pi$). Here, we have taken $G_v = 0$.}
\end{figure}
In FIG. \ref{fig:uds_fraction}, we have plotted the densities of different species for charge neutral quark matter as a function of 
baryon density for different values of the scaled axion field parameter $\theta = \langle a \rangle / f_a$. FIG. \ref{fig:uds_fraction} 
(a) corresponds to the case of $\theta = 0$ while FIG. \ref{fig:uds_fraction} (b) corresponds to the case of $\theta = \pi$. 
For $\theta = 0$, as $\mu_{\rm Q}$ ($\equiv \mu_B/3$) is increased from $\mu_{\rm Q} = 0$, the condensates remain constant with their vacuum expectation values and the baryon number density vanishes until $\mu_{\rm Q}^c(\theta = 0) = 367.37\ {\rm MeV}$. At this critical value of quark chemical potential there is a first order chiral transition with a jump in the baryon density from $n_{\rm B} = 0$ to $n_{\rm B} = 0.38\ {\rm fm}^{-3}$. The masses of up and down quarks also jumps from their vacuum value of $M_{u,d} = 367.65\ {\rm MeV}$ to $M_u = 81.9\ {\rm MeV}$, $M_d = 62.4\ {\rm MeV}$. Due to determinant interaction this chiral transition for light quark leads to a sharp decrease in a strange quark mass from its vacuum value $M_s = 549\ {\rm MeV}$ to $M_s = 465.79\ {\rm MeV}$. This leads to non-vanishing values for up, and down quarks densities as $n_u = 0.13\ {\rm fm}^{-3}$, and $n_d \sim 2 n_u = 0.254\ {\rm fm}^{-3} $ with a negligible density for the electron, $n_e = 0.0017\ {\rm fm}^{-3}$. In FIG. \ref{fig:uds_fraction} (b), these densities are shown as in FIG. \ref{fig:uds_fraction} (a) for the case of $\theta = \pi$. The general behavior of the density of each species is similar as is the case for $\theta = 0$ except that the chiral transition for $\theta = \pi$ takes place for a little smaller value of $\mu_{\rm Q}$ i.e. $\mu_{\rm Q}^c(\theta = \pi) = 351.4\ {\rm MeV}$.

\section{Equation of state for hadronic matter within RMF model \label{eos_hm_rmf_model}}
We shall consider here \ac{rmf} model to construct the \ac{eos} of neutron star matter (NSM) in the hadronic phase. In this framework, the interaction of baryons is realized through the exchange of various mesons. Here, we shall generalize the mean field model as considered in Ref. \cite{Kumar:2021hzo} for $npe\mu$ matter to include the lowest lying octet of baryons, ${\rm B} \equiv (n,\ p,\ \Lambda^0,\ \Sigma^{-,0,+},\ \Xi^{-,0})$, interacting with various mesons such as isoscalar-scalar $(\sigma)$, isoscalar-vector ($\omega$), isovector-vector ($\bm \rho$) mesons as well as mesons with hidden strangeness such as isoscalar-vector meson $(\phi)$. The Lagrangian can be written as \cite{Mishra:2001py, Tolos:2017lgv}
\begin{IEEEeqnarray}{rCl}
{\cal L} = \sum_{{\rm b}\in{\rm B}}{\cal L}_{\rm b}^{\rm kin} + {\cal L}_{\rm M}^{\rm kin} + {\cal L}_{\rm BM} - V_{\rm NL} + {\cal L}_{\rm l}^{\rm kin},
\label{lqhd}
\end{IEEEeqnarray}
where, ${\cal L}_{\rm b}$ is the kinetic term for the baryons given as
\begin{IEEEeqnarray}{rCl}
{\cal L}_{\rm b}^{\rm kin} = \bar\psi_{\rm b}(i\gamma^\mu\partial_\mu-m_{\rm b})\psi_{\rm b}. \label{lb}
\end{IEEEeqnarray}
Similarly the kinetic term for the mesons is given by

\begin{IEEEeqnarray}{rCl}
{\cal L}_{\rm M}^{\rm kin} &=& \frac{1}{2} \left[ \partial_{\mu}\sigma\partial^{\mu}\sigma - m_{\sigma}^2\sigma^2 \right] - \frac{1}{4}\Omega_{\mu\nu}\Omega^{\mu\nu} + \frac{1}{2} m_{\omega}^2\omega^2 - \frac{1}{4}{\rm \bf R}_{\mu\nu}{\rm \bf R}^{\mu\nu} + \frac{1}{2}m_{\rho}^2{\bm \rho}_{\mu}{\bm \rho}^{\mu} - \frac{1}{4}{\Phi}_{\mu\nu}{\Phi}^{\mu\nu} + \frac{1}{2}m_{\phi}^2\phi_{\mu}\phi^{\mu}, \label{lmkin}
\end{IEEEeqnarray}
with $\Omega^{\mu\nu} = \partial^{\mu}\omega^{\nu} - \partial^{\nu}\omega^{\mu}$, ${\rm \bf R}^{\mu\nu} = \partial^{\mu}{\rm \bm \rho}^{\nu} - \partial^{\nu}{\rm \bm \rho}^{\mu}$ are the vector mesonic field strength tensor. ${\cal L}_{BM}$ is the Lagrangian describing the baryon meson interactions having the form
\begin{IEEEeqnarray}{rCl}
{\cal L}_{\rm BM} &=& - \sum_{{\rm b}\ \in\ {\rm B}}\bar\psi_{\rm b}\gamma_{\mu}\left(g_{\omega {\rm b}}\omega^{\mu} + g_{\rho {\rm b}}{\bm \tau}_{\rm b}\cdot {\bm \rho}^{\mu} + g_{\phi {\rm b}}\phi^{\mu}\right)\psi_{\rm b},
\end{IEEEeqnarray}
where, $\psi_{\rm b}$ and $m_{\rm b}$ correspond to the baryonic field and its bare mass respectively, $g_{\alpha {\rm b}}$ for $\alpha \in \sigma, \omega^{\mu}, {\bm \rho}, \phi^{\mu}$ are the coupling constants of the baryons with the mesons. Similarly $V_{\rm NL}$ describes the nonlinear interaction of mesons and is given by
\begin{IEEEeqnarray}{rCl}
V_{\rm NL} = \frac{\kappa}{3!}(g_{\sigma {\rm N}}\sigma)^3 + \frac{\lambda}{4!}(g_{\sigma {\rm N}}\sigma)^4 - \Lambda_\omega g^2_{\rho {\rm N}} g^2_{\omega {\rm N}}(\omega_{\mu}{\bm \rho}^{\mu})^2 - \frac{\xi_\omega}{4!}(g_{\omega {\rm N}}^2\omega_{\mu}\omega^{\mu})^2.
\end{IEEEeqnarray}
Thus the nonlinear interaction mesons consists of (i) cubic and quartic self interaction of $\sigma$, (ii) a quartic mixing interaction between $\omega$ and ${\bm \rho}$ mesons and (iii) the quartic self interaction of $\omega$ mesons. This completes the description of hadronic part of the Lagrangian Eq. (\ref{lqhd}). However, to consider $\beta$-equilibrated charge neutral matter, one has to include leptons, ${\rm L} \equiv (e,\ \mu)$, with the corresponding Lagrangian being given as
\begin{IEEEeqnarray}{rCl}
{\cal L}_{\rm l}^{\rm kin} = \sum_{{\rm l} \in L}\bar\psi_{\rm l}(i\gamma^{\mu}\partial_{\mu} - m_{\rm l}) \psi_{\rm l}.
\end{IEEEeqnarray}
We shall be using the mean field approximation for the meson fields. This amounts to taking meson fields as classical fields while retaining the quantum nature for the baryonic fields. For uniform and static matter within this approximation, only the time like components of the vector fields and the isospin 3 component of the ${\bm \rho}$ field have non-vanishing values. The meson mean fields are thus denoted by $\sigma_0$, $\omega_0$, $\rho_{03}$, $\phi_0$. With the Lagrangian given in Eq. (\ref{lqhd}), one can identify the effective masses of the baryons in the mean field approximation as
\begin{IEEEeqnarray}{rCl}
m_{\rm b}^* = m_{\rm b} - g_{\sigma {\rm b}}\sigma_0,
\end{IEEEeqnarray}
and the effective chemical potentials $\mu_{\rm b}^*$ as
\begin{IEEEeqnarray}{rCl}
\mu_{\rm b}^* = \mu_{\rm b} - g_{\omega {\rm b}}\omega_0 - g_{\rho {\rm b}}I_{\rm 3b}\rho_{03} - g_{\phi {\rm b}}\phi_0.
\label{effmu}
\end{IEEEeqnarray}
From Eq. (\ref{effmu}) one can define the Fermi-momentum of each species as $k_{\rm Fb} = \sqrt{\mu_{\rm b}^*{}^2 - m_{\rm b}^{*}{}^2}$, for $\mu_{\rm b}{}^* > m_{\rm b}^{*}$ and zero otherwise. Thus the threshold condition for the appearance of the baryon of type b is given as 
\begin{IEEEeqnarray}{rCl}
\mu_{\rm b}^* \equiv \mu_{\rm b} - g_{\omega {\rm b}}\omega_0 - g_{\rho {\rm b}}I_{\rm 3b}\rho_{03} - g_{\phi {\rm b}}\phi_0 \geq m_{\rm b} - g_{\sigma {\rm b}}\sigma_0.
\label{threshold_hm_par}
\end{IEEEeqnarray}
For the Lagrangian, Eq. (\ref{lqhd}), within the mean field approximation, one can find the mesonic equations of motion given as
\begin{IEEEeqnarray}{rCl}
m_{\sigma}^2 \sigma_0 + \frac{\kappa}{2} g_{\sigma {\rm N}}^3\sigma^2 + \frac{\lambda}{6} g_{\sigma {\rm N}}^4\sigma^3 &=& \sum_{\rm b} g_{\sigma {\rm b}} n_{\rm b}^s, \label{fieldeqns.sigma}
\\
m_{\omega}^2 \omega_0 + 2 \Lambda_{\omega} g^2_{\rho {\rm N}} g^2_{\omega {\rm N}} \rho_{03}^2 \omega_{0} + \frac{\xi_\omega}{6} g_{\omega {\rm N}}^4\omega_{0}^3 &=& \sum_{\rm b} g_{\omega {\rm b}}n_{\rm b}, \label{fieldeqns.omega}
\\
m_{\rho}^2 \rho_{03} + 2 \Lambda_{\omega} g^2_{\rho {\rm N}} g^2_{\omega {\rm N}} \rho_{03} \omega_{0}^2 &=& \sum_{\rm b} g_{\rho {\rm b}}I_{3{\rm b}}n_{\rm b}, \label{fieldeqns.rho} 
\\
m_{\phi}^2 \phi_0 &=& \sum_{\rm b} g_{\phi {\rm b}} n_{\rm b}, \label{fieldeqns.phi}
\end{IEEEeqnarray}
and the energy density in the hadronic phase, $\epsilon_{\rm HP}$, given as
\begin{IEEEeqnarray}{rCl}
\epsilon_{\rm HP} &=& \frac{1}{\pi^2}\sum_{\rm b}k_{\rm Fb}^4 H\left(\frac{m_{\rm b}^*}{k_{\rm Fb}}\right) + \sum_{{\rm l}=e,\mu}\frac{1}{\pi^2} k_{\rm Fl}^4 H\left(\frac{m_{\rm l}^*}{k_{\rm Fl}}\right) + \frac{1}{2}m_{\sigma}^2\sigma_0^2 + V(\sigma_0) + \frac{1}{2} m_{\omega}^2\omega_0^2 + \frac{1}{2} m_{\rho}^2{\rho_{03}}^2, \label{energy.density.nm}
\end{IEEEeqnarray}
where, the function $H(z)$ is defined earlier in Eq. (\ref{hz}). The corresponding pressure, $p_{\rm HP}$, can be found using the thermodynamic relation as
\begin{IEEEeqnarray}{rCl}
p_{\rm HP} &=& \sum_{i={\rm B, L}} \mu_i n_i - \epsilon_{\rm HP}. \label{pressure.nm}
\end{IEEEeqnarray}
Further, we have also defined the baryonic number density 
\begin{IEEEeqnarray}{rCl}
n_{\rm b} = \langle {\psi}^{\dagger}_{\rm b}\psi_{\rm b} \rangle = \int_0^{k_{\rm Fb}} d{\zbf k} = \frac{k_{\rm Fb}^3}{3\pi^2}, \label{anb}
\end{IEEEeqnarray}
and the scalar baryon density 
\begin{IEEEeqnarray}{rCl}
n_{\rm b}^s = \langle \bar{\psi}_{\rm b}\psi_{\rm b} \rangle = \int \frac{m_{\rm b}^*}{E(k)} \Theta(k_{\rm Fb} - k) d{\zbf k} =  \frac{m_{\rm b}^* k_{\rm Fb}^2}{4\pi^2} G\left(\frac{m_{\rm b}^*}{k_{\rm Fb}}\right), \label{anbs}
\end{IEEEeqnarray}
\noindent where the function $G(z)$ is also already defined in Eq. (\ref{gz}) and $E(k) = \sqrt{m_{\rm b}^{*2}+ k^2}$ and $k_{\rm Fl}$ is the leptonic Fermi momenta i.e. $k_{\rm Fl}=\sqrt{\mu_{\rm l}^2-m_{\rm l}^2}$. 

To obtain the \ac{eos} for baryonic matter, we need to specify the meson-baryon couplings. The meson-nucleon couplings are chosen to satisfy saturation properties of nuclear matter. This parameter set NL3$\omega\rho$ is taken from Ref. \cite{Horowitz:2000xj}. One can generate the meson-hyperon vector couplings using SU(3) flavor quark counting rule as \cite{Schaffner:1993qj}
\begin{IEEEeqnarray}{rCl}
g_{\omega\Lambda} &=& g_{\omega\Sigma} = 2g_{\omega\Xi} = \frac{2}{3}g_{\omega N} \label{vec_hyp_couplings_omega} \\
g_{\rho\Lambda} &=& 0, \quad g_{\rho\Sigma} = 2g_{\rho\Xi} = \frac{2}{3}g_{\omega N} \label{vec_hyp_couplings_rho} \\
g_{\phi N} &=& 0, \quad 2 g_{\phi\Lambda} = 2 g_{\phi\Sigma} = g_{\phi\Xi} = \frac{2\sqrt{2}}{3}g_{\omega N}. \label{vec_hyp_couplings_phi}
\end{IEEEeqnarray}
The scalar couplings of mesons with hyperons, on the other hand, are fixed from the potential depth of hyperons in nuclear matter \cite{Schaffner:1993qj, Tolos:2017lgv, Mishra:2001py}. The hyperonic potential in the nuclear matter satisfy the equation, 
\begin{IEEEeqnarray}{rCl}
    U_{Y}^{N} = -g_{\sigma Y}\sigma_0 + g_{\omega Y}\omega_0.
\end{IEEEeqnarray}
The coupling of hyperons to sigma field is adjusted to reproduce the hyperon potentials in strange hadronic matter is taken as $U_{\Lambda}^{N} = -28\ {\rm MeV}$, $U_{\Sigma}^{N} = 30\ {\rm MeV}$, and $U_{\Xi}^{N} = -18\ {\rm MeV}$ at saturation density \cite{Schaffner:1993qj, Tolos:2017lgv, Mishra:2001py}. However, it may be noted that the precise value of $U_{\Xi}^{N}$ at saturation is not well constrained though it is known to be attractive. These values of hyperonic potentials lead to the couplings $g_{\sigma\Lambda} = 0.6113\ g_{\sigma N}$, $g_{\sigma\Sigma} = 0.4665\ g_{\sigma N}$, and $g_{\sigma\Xi} = 0.3157\ g_{\sigma N}$. The complete list of parameters for the NL3$\omega\rho$ model that we use for the description of hadronic matter is given in TABLE \ref{table.nl3wr.parameters}.
\begin{table}[b]
\caption{The complete list of parameters for NL3$\omega\rho$ model \cite{Horowitz:2000xj}. \label{table.nl3wr.parameters}}
\begin{tabular}{|cr||cr|}
\hline\hline
{\bf Masses}       & {\bf Values}   &  {\bf Meson}            & {\bf Values}  \\
                   &   ({\bf MeV})  &  {\bf Couplings}        &               \\
\hline
{$m_{\rm N}$}      & 939            &  $m_{\sigma}$           & 508.194 (MeV) \\
{$m_{\Lambda}$}    & 1115.683       &  $m_{\omega}$           & 782.501 (MeV) \\
{$m_{\Sigma^{-}}$} & 1197.449       &  $m_{\rho}$             & 763     (MeV) \\
{$m_{\Sigma^{0}}$} & 1192.624       &  $m_{\phi}$             & 1020    (MeV) \\
{$m_{\Sigma^{+}}$} & 1189.37        &  $g_{\sigma {\rm N}}^2$ & 104.3871      \\
{$m_{\Xi^{-}}$}    & 1314.86        &  $g_{\omega {\rm N}}^2$ & 165.5854      \\
{$m_{\Xi^{0}}$}    & 1321.71        &  $g_{\rho {\rm N}}^2$   & 79.6          \\
                   &                &  $\kappa$               & 3.8599 (fm$^{-1}$) \\
                   &                &  $\lambda$              & -0.015905 \\
                   &                &  $\Lambda_{\omega}$     & 0.03 \\
\hline\hline
\end{tabular}%
\end{table}%

For the consideration of hadronic matter in the context of \ac{ns}, we also have to take into account that the matter in the \ac{ns} core to be $\beta$ equilibrated and globally charge neutral. Consequently the chemical potentials $\mu_{\rm b}$ and the particle density $n_{\rm b}$ satisfy the following conditions, 
\begin{IEEEeqnarray}{rCl}
\mu_{\rm b} &=& b_{\rm b} \mu_{\rm B} - q_{\rm b}\mu_{\rm E}, \label{beta_equalibrium_baryon}
\\
\sum_{{\rm i} = {\rm B, L}} q_{\rm i} n_{\rm i} &=& 0, \label{charge_neutrality_baryon}
\end{IEEEeqnarray}
where, $\mu_{\rm B}$ is the baryon chemical potential while $b_{\rm b}$ and $q_{\rm b}$ are the baryon number and electric charge of the corresponding baryon. The Eqs. (\ref{fieldeqns.sigma} - \ref{fieldeqns.phi}) along with the constraints Eqs. (\ref{beta_equalibrium_baryon} - \ref{charge_neutrality_baryon}) comprise a coupled set of equations which are solved self-consistently to determine the mesonic mean fields and the electric chemical potential, $\mu_{\rm E}$ for a given $\mu_{\rm B}$. These are used in Eqs. (\ref{energy.density.nm} and \ref{pressure.nm}) to get \ac{eos} of hadronic component of \ac{nsm}.

\begin{figure}
     \begin{subfigure}[t]{0.60\textwidth}
     \includegraphics[width=\textwidth]{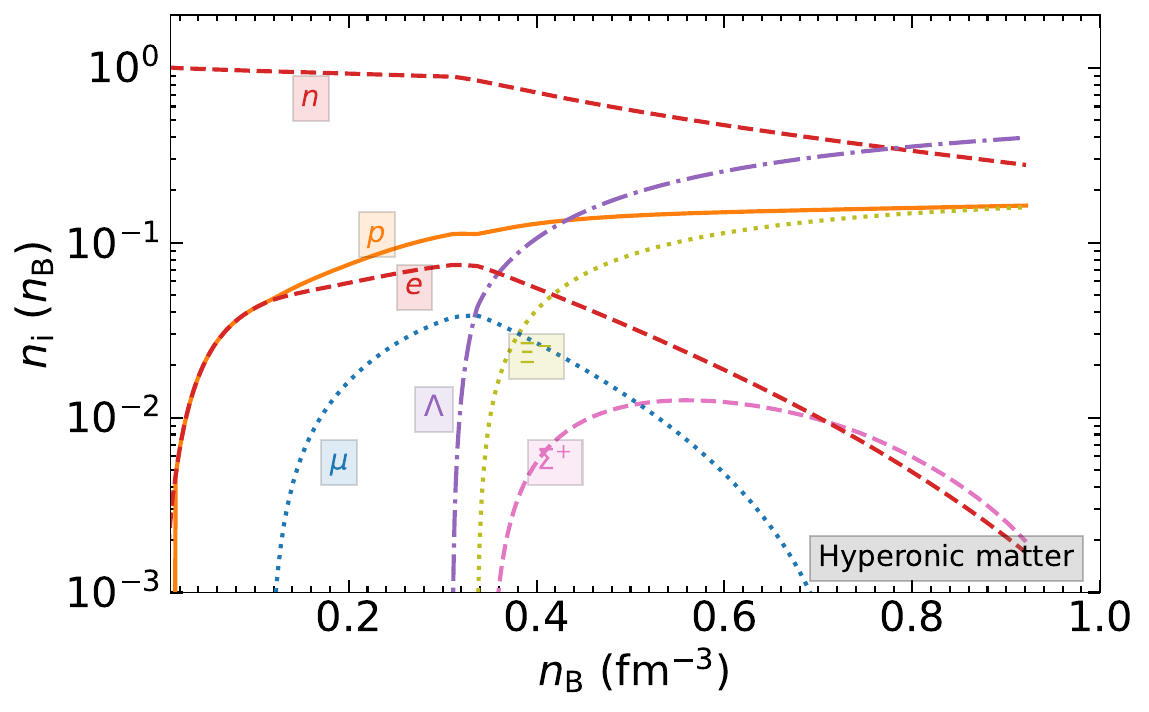}
     \end{subfigure}
\caption{\label{fig:hadronic_fractions} Population densities of different species for charge neutral hypronic matter as a function of baryon density.}
\end{figure}

In FIG. \ref{fig:hadronic_fractions}, we display the composition of charge neutral  hadronic matter i.e. the densities of different species as a function of baryon number density using the parameter set as given in TABLE \ref{table.nl3wr.parameters}. At $n_{\rm b} \sim 2n_0 = 0.31\ {\rm fm}^{-3}$, $\Lambda$, the lightest hyperon appears. It is also observed that for a little higher value of density, the negatively charged hyperon $\Xi^{-}$ appears which competes with the leptons to maintain charge neutrality. Although the vacuum mass of $\Sigma^{+}$ is smaller than the same for $\Xi^{-}$, the threshold density for the appearance of $\Xi^{-}$ is smaller than that of $\Sigma^{+}$. This is because the repulsion due to $\rho$ meson for $\Xi^{-}$ is smaller then that for $\Sigma^{+}$ which leads to a value of $\mu^*_{\Xi^{-}}$ large enough to have the threshold density for the occurrence of $\Xi^{-}$ earlier as compared to $\Sigma^{+}$. We might note here that this is in line with the extrapolated $\Sigma^{+}$ atomic data \cite{Mares:1995bm}, which suggest that $\Sigma^{+}$ may feel repulsion at high density, which would mitigate against its appearance in dense matter \cite{Glendenning:2000jx}. The appearance of $\Xi^{-}$ leads to a depletion of lepton densities consistent with charge neutrality condition seen as seen in FIG. \ref{fig:hadronic_fractions}. As may be observed, the proton density increases with density and saturates at higher densities. The appearance of $\Sigma^{+}$ maintains the charge neutrality condition. We might note here the threshold condition for a given species to appear is given by Eq. (\ref{threshold_hm_par}). Thus, the order in which the hyperons appear with increasing density is rather sensitive to hyperon-meson coupling constants. With the couplings chosen in the present work, the threshold density of $\Sigma^{+}$ is smaller than $\Sigma^{0}$, and $\Sigma^{-}$. This is due to two reasons. First of all the vacuum mass of $\Sigma^{+}$ is smaller than that of $\Sigma^{0}$, and $\Sigma^{0}$ and secondly its couplings to the isovector-vector $\rho$ meson decided by the isospin leads to a higher effective chemical potential compared to $\Sigma^{0}$ and $\Sigma^{-}$. Further, it may also be observed that after an initial rise of $\Sigma^{+}$ density it starts decreasing at a higher density when density of $\Xi^{-}$ saturates. This is because electron density keeps decreasing and to maintain charge neutrality, $\Sigma^{+}$ also decreases at higher densities.

\section{Hadron-quark phase transition in presence of axion \label{hqpt}}
The baryon number density at which the \ac{hqpt} occurs is not known precisely from the first principle lattice \ac{qcd} simulation. However, it is expected from various model calculations that such a transition occurs at a density which is a few times nuclear matter saturation density. There are two limiting approaches to study the phase transition between hadronic matter and quark matter. One is through the Maxwell construction, where a sharp first order phase transition with a local charge conservation is considered. The other is the Gibbs construction, where a mixed phase exists and the charge is conserved globally. The crucial quantity which decides which type of construction is followed by \ac{hqpt} is the value of surface tension in the hadron-quark interface. For the large values of surface tension one can have a Maxwell construction while for small values one can have a Gibbs construction for the \ac{hqpt}. As the precise value of surface tension is not known both the scenarios are plausible. In this work, we use the Gibbs construction for \ac{hqpt} which has been nicely outlined in Ref. \cite{Schertler:1999xn, Kumar:2021hzo}. Here, the charge neutrality is achieved with a positively charged hadronic matter mixed with a negatively charged quark matter in an appropriate amount so that the global charge neutrality is maintained. Here, the pressures of both the phases are the functions of two independent chemical potentials $\mu_{\rm B} (=3\mu_{\rm Q})$ and $\mu_{\rm E}$. The Gibbs condition for the equilibrium at zero temperatures is given as 
\begin{IEEEeqnarray}{rCl}
    p_{\rm HP} (\mu_{\rm B}^c, \mu_{\rm E}^c) = p_{\rm QP} (\mu_{\rm B}^c, \mu_{\rm E}^c) = p_{\rm MP} (\mu_{\rm B}^c, \mu_{\rm E}^c), \label{gibbs_condition}
\end{IEEEeqnarray}
where, HP, QP and MP represent hadronic phase, quark phase and mixed phase respectively and $\mu_{\rm B}^c$ and $\mu_{\rm E}^c$ are the critical baryon and electric chemical potentials at which the \ac{hqpt} occurs. 
The global charge neutrality condition is defined through the equation, 
\begin{IEEEeqnarray}{rCl}
    \chi q_{\rm QP} + (1 - \chi)q_{\rm HP} = 0, \label{global_charge_neutrality}
\end{IEEEeqnarray}
where, $q_{\rm HP}$ and $q_{\rm QP}$ are the total charge densities in HP and QP respectively. The parameter $\chi$ is the volume fraction of quark matter in mixed phase given as
\begin{IEEEeqnarray}{rCl}
\chi = \frac{V_{\rm QP}}{V_{\rm QP} + V_{\rm HP}}.
\end{IEEEeqnarray}
For a given $\mu_{\rm B}$, we calculate the electric chemical potential $\mu_{\rm E}$ such that the pressures in both the phases are equal and satisfy Eq. (\ref{gibbs_condition}). Further, imposing the globally charge neutrality condition, Eq. (\ref{global_charge_neutrality}), one obtains the volume fraction $\chi$ occupied by quark matter in mixed phase. Along with that we can define the energy density and baryon number density in the mixed phase as 
\begin{IEEEeqnarray}{rCl}
    \epsilon_{\rm MP} &=& \chi \epsilon_{\rm QP} + (1 - \chi)\epsilon_{\rm HP} \label{energy_density_mp} \\
    \rho_{\rm MP} &=& \chi \rho_{\rm QP} + (1 - \chi) \rho_{\rm HP}. \label{baryon_density_qp}
\end{IEEEeqnarray}

\begin{figure}
     \begin{subfigure}[t]{0.49\textwidth}
      \includegraphics[width=\textwidth]{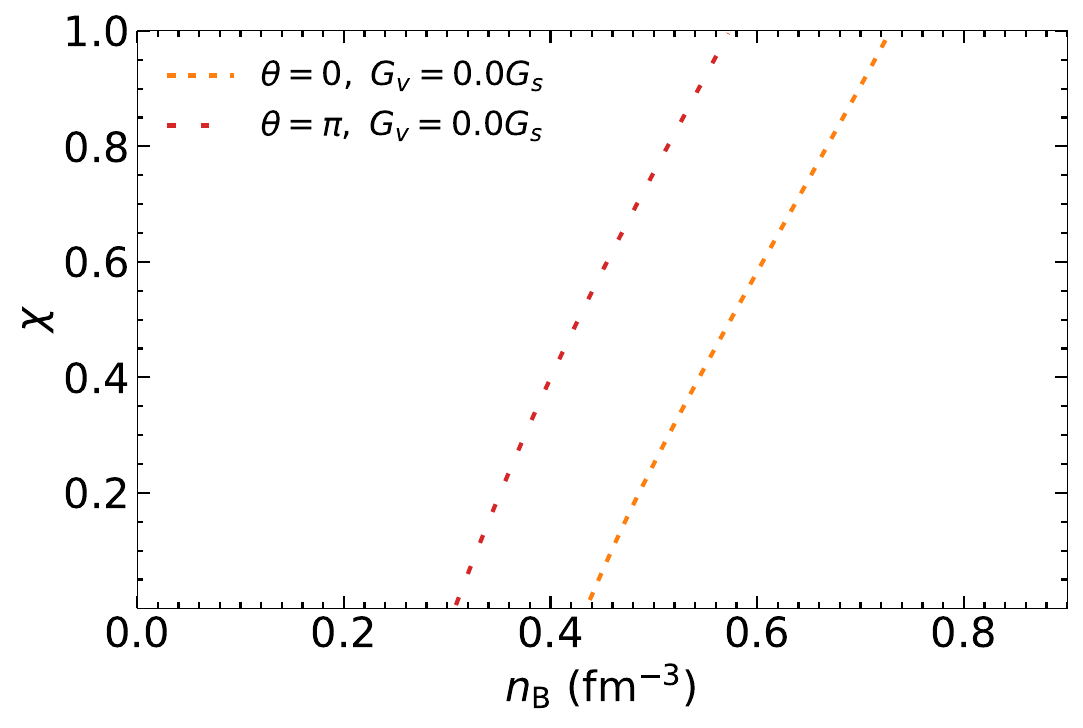}
      \caption{}
    \end{subfigure}
     \begin{subfigure}[t]{0.49\textwidth}
      \includegraphics[width=\textwidth]{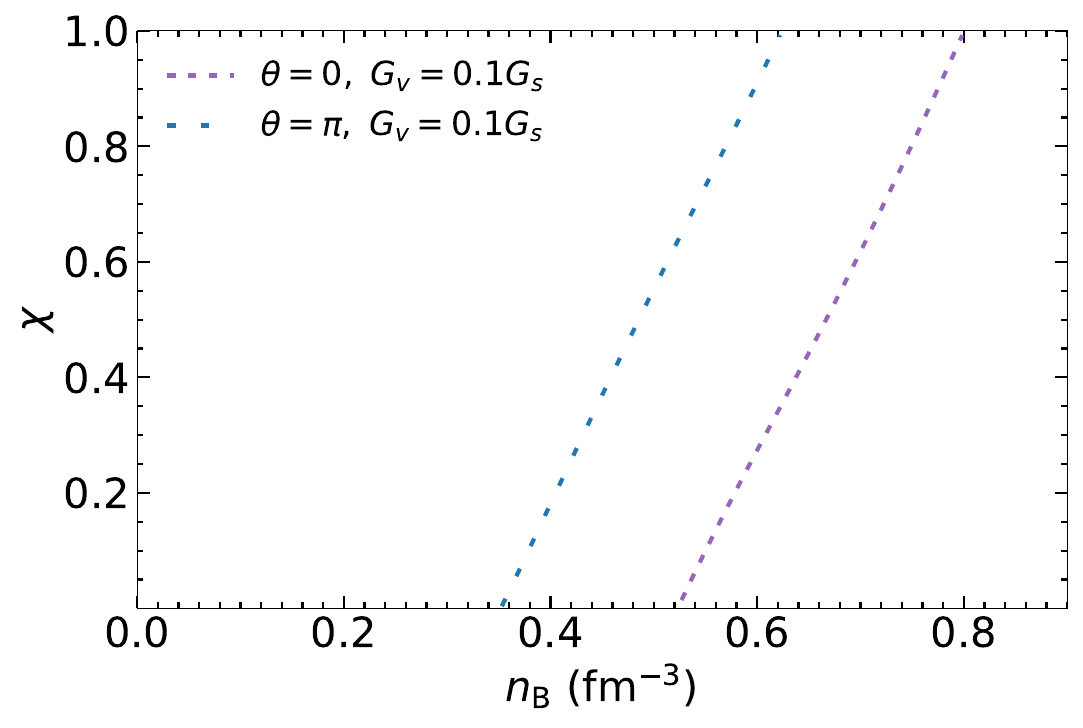}
      \caption{}
    \end{subfigure}
    \begin{subfigure}[t]{0.49\textwidth}
      \includegraphics[width=\textwidth]{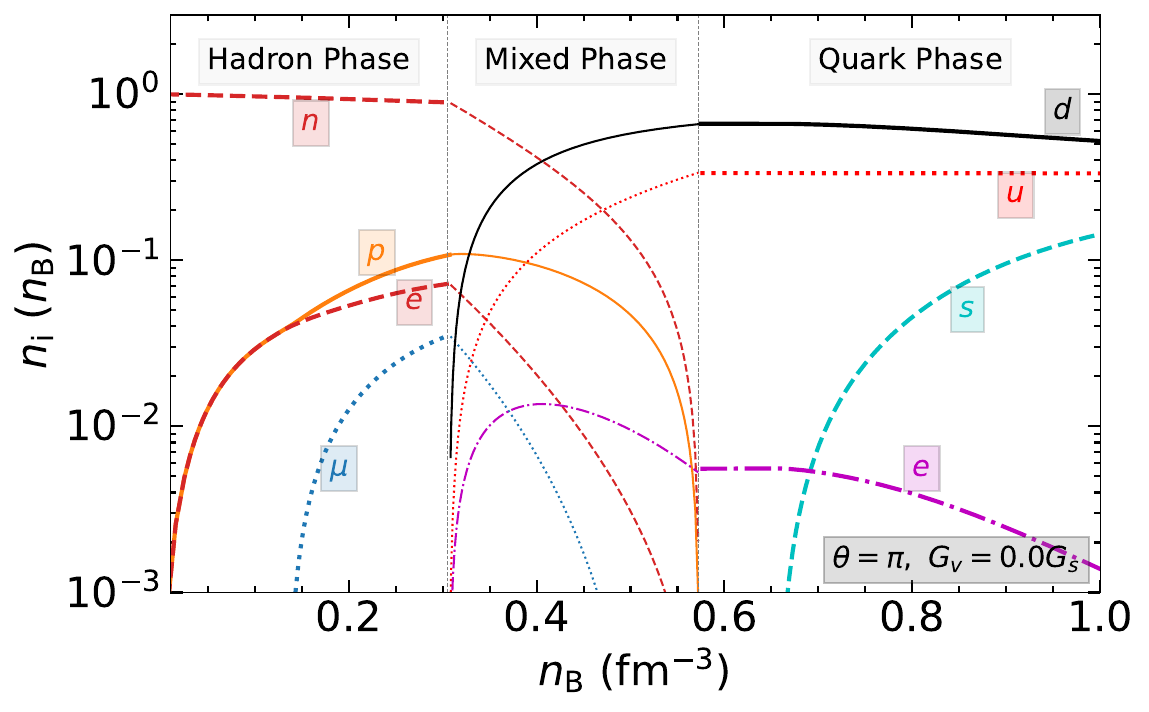}
      \caption{}
    \end{subfigure}
    \begin{subfigure}[t]{0.49\textwidth}
      \includegraphics[width=\textwidth]{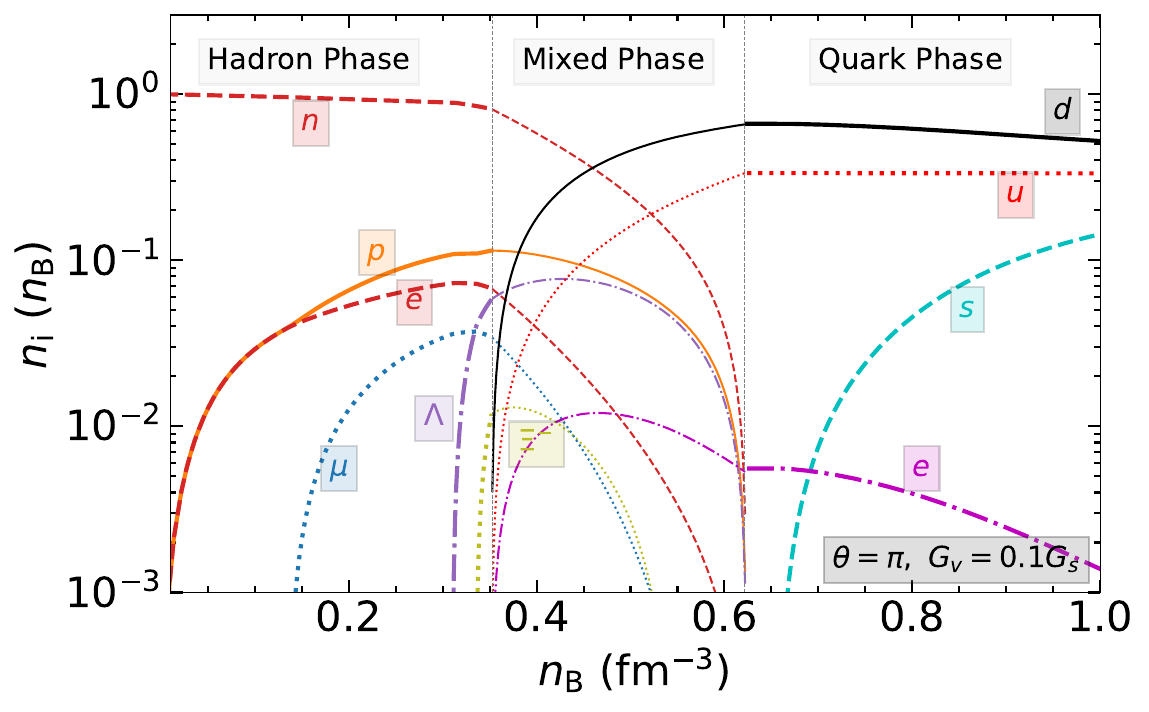}
      \caption{}
    \end{subfigure}
\caption{\label{fig:particle_fractions_and_chi} Upper panel: Quark matter fraction as a function of baryon density in Gibbs construction for the mixed phase for different $\theta$ at constant $G_v$. Left(a) for $G_v = 0$ and $\theta = 0,\ \pi$, Right(b) for $G_v = 0.1G_s$ and $\theta = 0,\ \pi$. Lower panel: The particle densities in the different phases (hadron phase, mixed phase, quark phase) when $\theta = \pi$ and $G_v = 0.0$ Left(c), $G_v = 0.1 G_s$ Right(d). The two vertical gray dotted lines in the bottom depict the beginning and end of the mixed phase.}
\end{figure}
In FIG. \ref{fig:particle_fractions_and_chi} we show some characteristics of the mixed phase structure for different values of scaled axion field $\theta$ as well as different values of vector coupling $G_v$ in the \ac{njl} model. In FIG. \ref{fig:particle_fractions_and_chi} (a), for $G_v = 0$, we have plotted the quark fraction as a function of baryon density in the mixed phase for $\theta = 0$ (orange dashed curve) and $\theta = \pi$ (red dashed curve). For $\theta = 0$, the mixed phase starts at $n_{\rm B} = 0.44\ {\rm fm}^{-3}$ upto which $\chi = 0$. As the density increases the quark fraction increases. It reaches to its maximum value $\chi = 1$ at $n_{\rm B} = 0.74\ {\rm fm}^{-3}$. This also signals the end of the mixed phase. As the density increases further, the system is in pure quark matter phase. For $\theta = \pi$, the threshold for the mixed phase decreases to $n_{\rm B} = 0.31\ {\rm fm}^{-3}$. As density increases the quark matter fraction $\chi$ increases to its maximum value of $\chi = 1$ at $n_{\rm B} = 0.57\ {\rm fm}^{-3}$. In FIG. \ref{fig:particle_fractions_and_chi} (b), we plot the same curves but with $G_v = 0.1G_s$. The inclusion of the repulsive interaction makes \ac{eos} for quark matter stiffer and the threshold for mixed phase for $\theta = 0$ becomes $n_{\rm B} = 0.53\ {\rm fm}^{-3}$ which is larger compared to the same for $G_v = 0$ case. In this case $\chi$ reaches to its maximum value of unity at $n_{\rm B} = 0.79\ {\rm fm}^{-3}$. For $\theta = \pi$ the threshold of the mixed phase to occur reduces to $n_{\rm B} = 0.35\ {\rm fm}^{-3}$ and becomes one at $n_{\rm B} = 0.62\ {\rm fm}^{-3}$. Thus the behavior of repulsive interaction $G_v$ and scaled axion field $\theta$ act in a complementary manner regarding the appearance of quark matter phase. Inclusion of axions makes the chiral phase transition earlier while presence of $G_v$ makes the chiral transition later as the density increases. This will have interesting consequences regarding the stability of hybrid \ac{ns} as has been observed in Ref. \cite{Lopes:2022efy}. We next show the particle content for the charge neutral matter in FIG. \ref{fig:particle_fractions_and_chi}, for $\theta = \pi$ where the axion effects play a larger role for two different values of the vector couplings $G_v = 0$ and $G_v = 0.1G_s$. As can be observed in FIG. \ref{fig:particle_fractions_and_chi}(c), for $G_v = 0$, the mixed phase starts at $n_{\rm B} = 0.31\ {\rm fm}^{-3}$. In the hadronic phase the neutron density dominates with a smaller fraction of proton and leptons appearing to have the charge neutral hadronic matter. At $n_{\rm B} = 0.31\ {\rm fm}^{-3}$, the mixed phase starts and nucleon fraction decreases while quark fraction starts increasing. At $n_{\rm B} = 0.57\ {\rm fm}^{-3}$, a pure quark phase takes over with the down quark density as twice of the up quarks to maintain the charge neutrality in quark matter phase. Let us note that the threshold for mixed phase to appear is just a little below the threshold for hyperon to appear in the hadronic phase. On the other hand for $G_v = 0.1G_s$ as shown in FIG. \ref{fig:particle_fractions_and_chi}(d), the threshold for mixed phase to appear is higher than the threshold for occurrence of hyperons $\Lambda$ and $\Xi^{-}$ in HP. Therefore, for $G_v = 0.1G_s$, it is possible to have a mixed phase with hyperonic matter and quark matter. Thus for $\theta = \pi$, it is possible to have a mixed phase with hyperonic matter {\em only} with a non-vanishing vector interaction in the \ac{njl} model. On the other hand, for $\theta = 0$, where chiral transition takes place at a little higher density, it is possible to have hyperons in the hadronic phase as well as in the mixed phase.

\section{Neutron star and its non-radial oscillation \label{neutron_star_and_non_radial_oscillation}}
The general static spherically-symmetric metric which describe the geometry of a static \ac{ns} can be written as
\begin{IEEEeqnarray}{rCl}
ds^2 &=& e^{2\nu(r)} dt^2-e^{2\lambda(r)} dr^2-r^2 (d\theta^2+\sin^2\theta d\phi^2), \label{metric}
\end{IEEEeqnarray}

where, $\nu(r)$ and $\lambda(r)$ are the metric functions. It is convenient to define the mass function, $m(r)$ in the favor of $\lambda$(r) as 
\begin{equation}
e^{2\lambda(r)} = \left(1-\frac{2m(r)}{r}\right)^{-1}.
\end{equation}
Starting from the line element, Eq. (\ref{metric}), one can obtain the equations governing the structure of spherical compact objects, the \ac{tov} equations, as 
\begin{IEEEeqnarray}{rCl}
\frac{dp(r)}{dr} &=& -\left(\epsilon +p \right)\frac{d\nu }{dr}, \label{tov.pressure}
\\
\frac{dm(r)}{dr} &=& 4\pi r^2 \epsilon, \label{tov.mass}
\end{IEEEeqnarray}
\begin{IEEEeqnarray}{rCl}
\frac{d\nu(r)}{dr} &=& \frac{m+4 \pi r^3p}{r(r-2m)}. \label{tov.phi}
\end{IEEEeqnarray}
In the above set of equations $\epsilon(r)$, $p(r)$ are the energy density and the pressure respectively. $m(r)$ is the mass of the compact star enclosed within a radius $r$. The boundary conditions $m(r=0) = 0$ and $p(r=0)=p_c$ and $p(r=R) = 0$, where $p_c$ is the central pressure lead to equilibrium configuration in combination with the \ac{eos} of \ac{ns} matter, thus obtaining radius $R$ and mass $M = m(R)$ of \ac{ns} for a given central pressure, $p_c$, or energy density, $\epsilon_c$. For a set of central energy densities $\epsilon_c$, one can obtain the mass-radius (M-R) curve. 

Using Einstein field equations and baryon number conservation, the theory for the \ac{nro}s of \ac{ns} was developed in Ref. \cite{1967ApJ...149..591T}. The perturbation of fluid in the star as described by the Lagrangian displacement vector $\xi^{\alpha}$ in terms of perturbing functions $Q(r,t)$ and $Z(r,t)$ is 
\begin{IEEEeqnarray}{rCl}
\xi^i = \left(e^{-\lambda(r)}Q(r,t),\ -Z(r,t)\partial_{\theta},\ 0 \right) r^{-2}P_{l}(\cos \theta).
\end{IEEEeqnarray}
We choose a harmonic time dependence for the perturbation functions, $Q(r,t)$ and $Z(r,t)$, which are proportional to $e^{-i\omega t}$, with `$\omega$' being the frequency. Further, we do not consider, here, toroidal deformation. The perturbing functions can be shown to satisfy in general the first order coupled differential equation within the Cowling approximation \cite{Kumar:2021hzo}. 

\begin{IEEEeqnarray} {rCl}
Q' - \frac{1}{c_e^2}\left[\omega^2 r^2e^{\lambda-2\nu}Z+\nu' Q\right]+l(l+1)e^\lambda Z = 0, \label{qprime_chap1}
\\
Z' - 2\nu' Z+e^\lambda \frac{Q}{r^2}-\nu'\left(\frac{1}{c_e^2}-\frac{1}{c_s^2}\right)\left(Z+\nu'e^{-\lambda+2\nu}\frac{Q}{\omega^2r^2}\right)=0 \label{zprime_chap1},
\end{IEEEeqnarray}
where, the prime denotes the radial derivative. For a detailed comprehensive derivation of equations, Eqs. (\ref{qprime_chap1} and \ref{zprime_chap1}), we refer \cite{Kumar:2021hzo}. In Eqs. (\ref{qprime_chap1} and \ref{zprime_chap1}), $c_e^2 = dp/d\epsilon = p^\prime/\epsilon^\prime$ is square of the equilibrium speed of sound which is the derivative of \ac{eos} in $\beta$-equilibrium. One the other hand, assuming the weak interaction time scales are slow compared to \ac{nro} time scale, the adiabatic sound speed is $c_s^2 = \left(\partial p/\partial \epsilon \right)_{y_i,s}$. The pre-factor of the last term on the left hand side of Eq. (\ref{zprime_chap1}) is proportional to the relativistic Brunt-V\"{a}is\"{a}la frequency \cite{McDermott:1983}. This term is responsible for the gravity mode ($g$ mode) oscillations \cite{Kumar:2021hzo}. In the present case, we shall ignore this term as we shall be discussing here $f$ mode oscillations only. The coupled first order differential equations for $Q(r)$ and $Z(r)$ given by Eqs. (\ref{qprime_chap1} and \ref{zprime_chap1}) are to be solves with appropriate boundary conditions n the center and the surface of the star.  Near the center of the compact stars, the behavior of the functions $Q(r)$ and $Z(r)$ are given by, \cite{Sotani:2010mx}
\begin{eqnarray}
Q(r) = Cr^{l+1} \quad \mathrm{and} \quad Z(r)=-Cr^l/l, \label{intital.conditions.of.w.and.v}
\end{eqnarray}
where, $C$ is an arbitrary constant and $l$ is the order of the oscillation. The other boundary condition is the vanishing of the Lagrangian perturbation pressure, $i.e.$ $\Delta p=0$. The vanishing of $\Delta p$ at surface leads to the boundary condition \cite{Kumar:2021hzo} 
\begin{IEEEeqnarray}{rCl}
\left[\omega^2 r^2e^{\lambda-2\nu}Z+\nu' Q\right]_{r=R} = 0. \label{surface_conditions}
\end{IEEEeqnarray}
There are extra conditions for the continuity conditions for $Q(r)$ and $Z(r)$ in case there is a discontinuity in the energy density as for example in Maxwell construction of phase transition. In the present case of Gibbs construct for the phase transition, the energy density as such is continuous. 

For a given central pressure, we solve the \ac{tov} equations, Eqs. (\ref{tov.pressure}) and (\ref{tov.mass}) to get the profile of the unperturbed metric functions $\nu(r)$ and $\lambda(r)$ as well as mass $m(r)$ as a function of radial distance from the center of the star. For a given frequency, we solve the pulsating equations, Eqs. (\ref{qprime_chap1}) and (\ref{zprime_chap1}). These solutions are substituted in to the left hand side of equation, Eq. (\ref{surface_conditions}). Then the value of $\omega$ is varied in such a way that the boundary condition, Eq. (\ref{surface_conditions}) has to be satisfied. This gives the frequency as a function of mass and radius. It should be noted that there can be multiple solutions of $\omega$ satisfying the boundary condition for different initial trial values of $\omega$. These different solutions for $\omega$ correspond to frequencies of different non-radial modes of oscillating compact star.

\section{results and discussions} \label{results_and_discussion}
As mentioned earlier, we shall consider \ac{njl} model with axion to describe quark matter while \ac{rmf} model to describe nuclear matter. Let us first discuss, \ac{cp} violation and chiral symmetry breaking  in quark matter at finite density. In subsection \ref{sub:a}, we give the results for \ac{cp} violation and chiral symmetry breaking in presence of axions in dense quark matter at zero temperature. In subsection \ref{sub:b}, we give the results for charge neutral matter and \ac{hqpt}. In subsection \ref{sub:c}, we use these findings and give the results for the mass-radius relations and \ac{ns}'s structure for various values of axion parameter, $\theta$ and corresponding $f$ mode oscillations. 

\subsection{CP violation and chiral symmetry breaking in dense quark matter} \label{sub:a}
For a given quark chemical potential, $\mu_{\rm Q}$, and axion parameter $\theta = {\langle a\rangle}/{f_a}$, we solve the mass gap equations, Eqs. (\ref{msu}-\ref{mps}) self-consistently. We shall present in the following the results, first, for the isospin symmetric quark matter so that $u$ and $d$ quark masses are equal i.e. $M_u = M_d$. Thus the equations, Eqs. (\ref{msu}-\ref{mps}) reduced to four coupled gap equations: two for the scalar condensates related to the two masses $M^u_s = M^d_s$, $M^s_s$ and two for pseudoscalar condensates related to the corresponding mass parameters $M^u_p = M^d_p$, $M^s_p$. When the charge neutrality condition is imposed the chemical potentials of $u$ and $d$ quarks are not the same and one needs to solve seven coupled equations: six for the mass components and one for the charge neutrality condition, Eq. (\ref{charge_neutrality_qm}). The solutions of these equations are substituted in the thermodynamic potential given in Eq. (\ref{pressure_qm}). It may be noted that near the chiral transition there are more than one solutions for the gap equations. We choose the solution which corresponds to the minimum thermodynamic potential or the maximum value of pressure. To discuss the general behavior of chiral and \ac{cp} transition in dense quark matter, we first discuss the case when $G_v=0$ and when charge neutrality conditions are not imposed so that all the three quarks have the same chemical potentials. 

\begin{figure}
    \begin{subfigure}[t]{0.49\textwidth}
      \includegraphics[width=\textwidth]{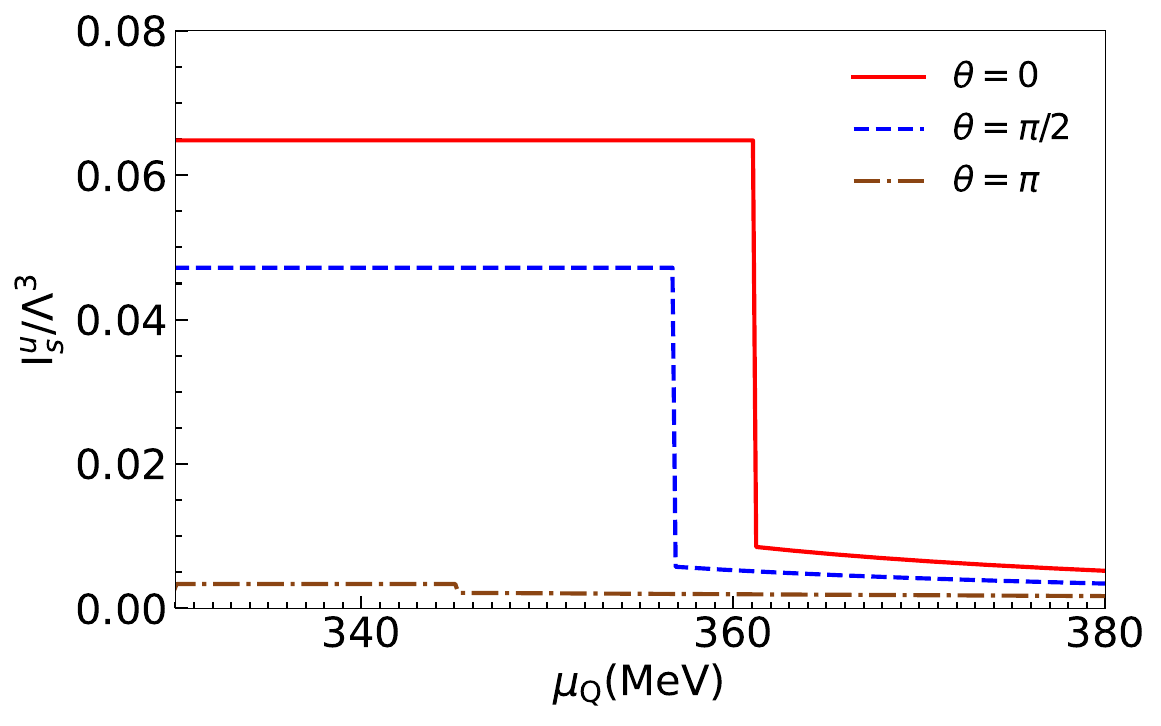}
      \caption{}
    \end{subfigure}
    \begin{subfigure}[t]{0.49\textwidth}
      \includegraphics[width=\textwidth]{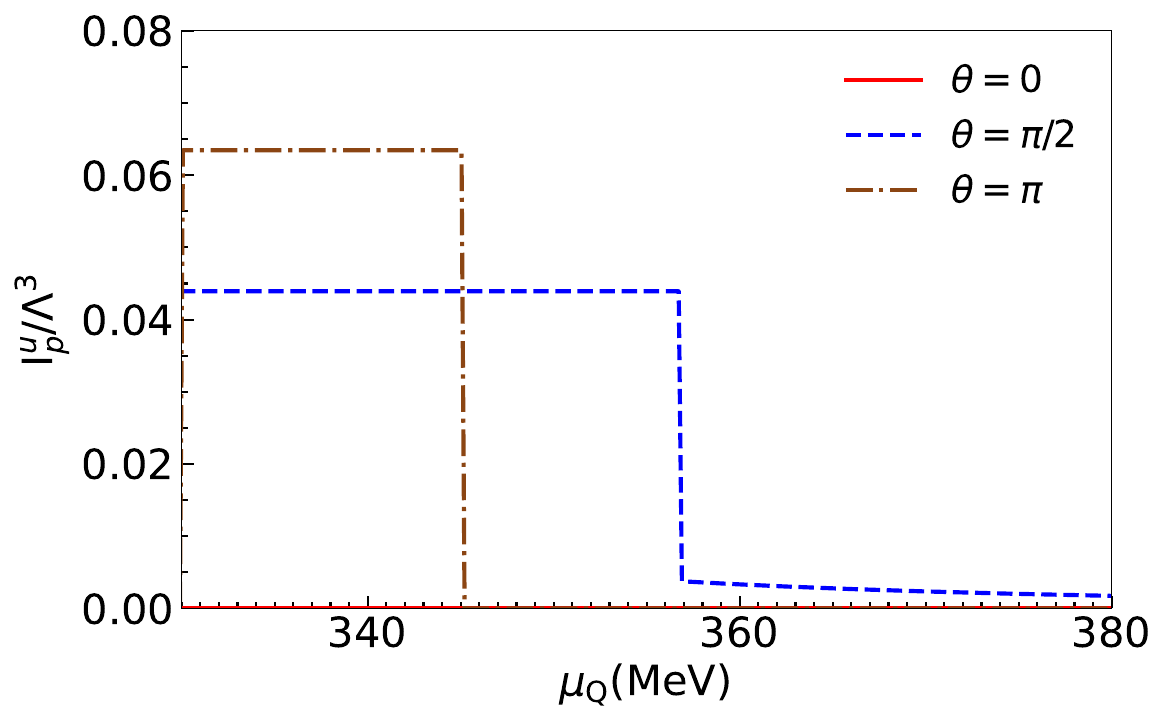}
      \caption{}
    \end{subfigure}
    \begin{subfigure}[t]{0.49\textwidth}
      \includegraphics[width=\textwidth]{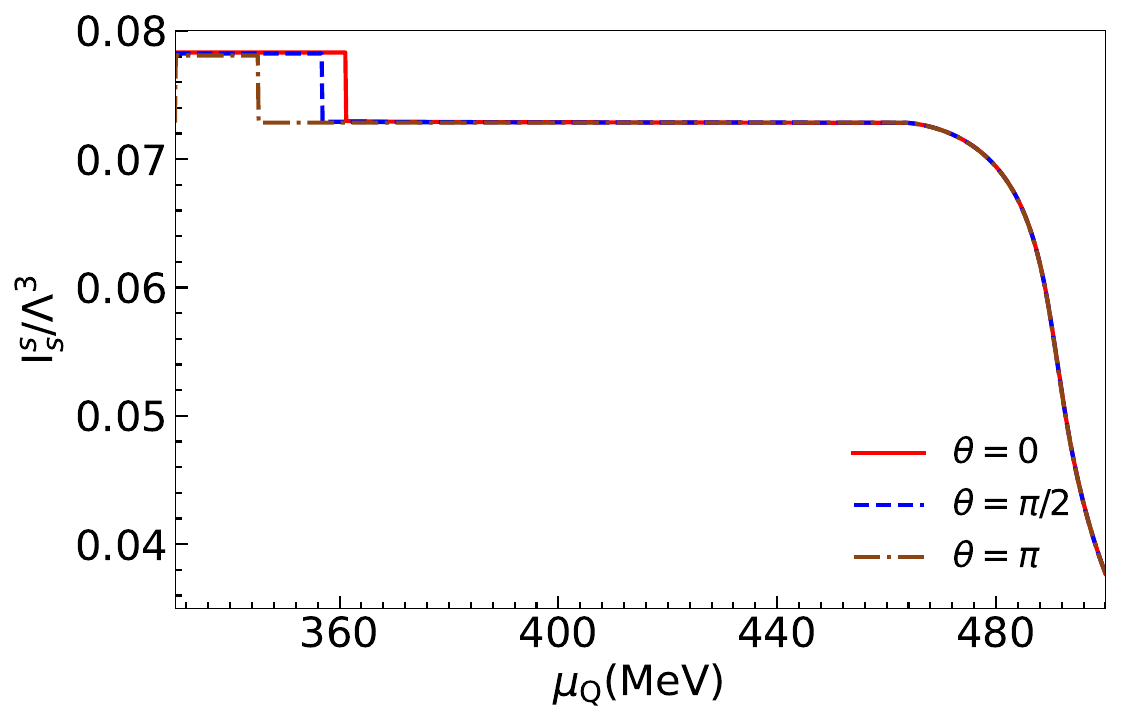}
      \caption{}
    \end{subfigure}
    \begin{subfigure}[t]{0.49\textwidth}
      \includegraphics[width=\textwidth]{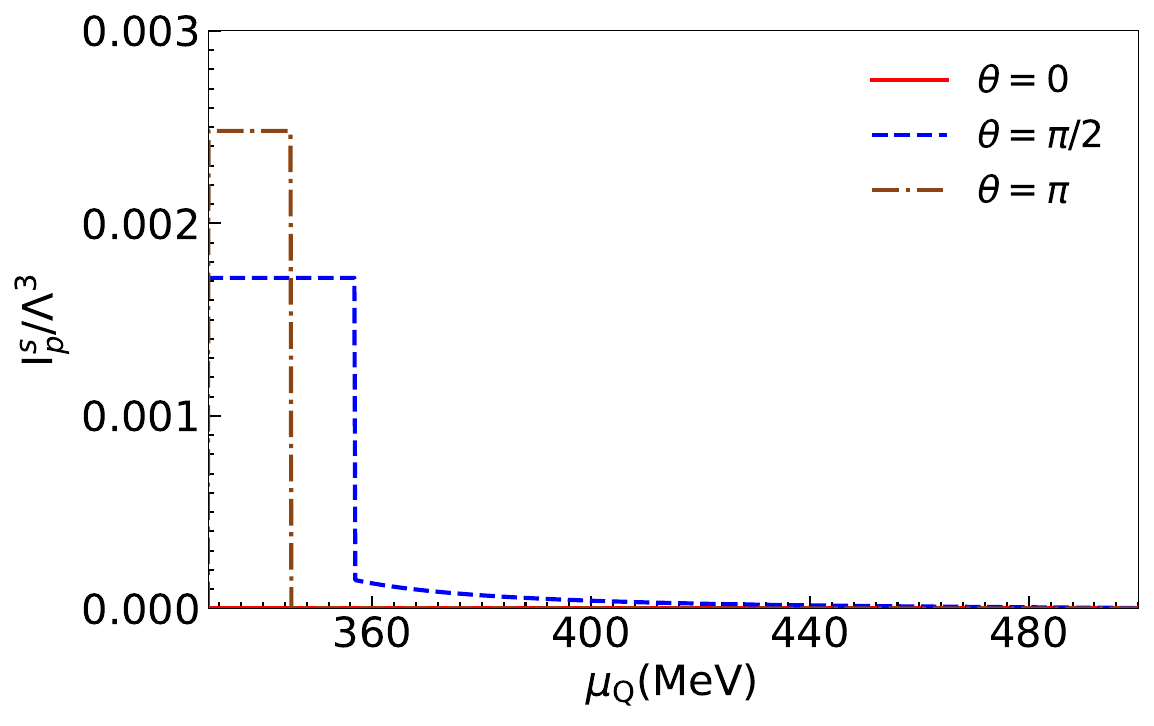}
      \caption{}
    \end{subfigure}
    \caption{\label{fig:condensates_amub} Upper panel: The variations of normalized scalar condensate, $I^u_s=\langle{\bar u u\rangle}/\Lambda^3$ (Fig. 4a) and pseudo-scalar condensate  $I^u_p=\langle{\bar u i\gamma^5u}\rangle/\Lambda^3$ (Fig. 4b) for the up quarks as a function of quark chemical potential $\mu_Q=\mu_B/3$ for $\theta = 0,\ \pi/2,\ \pi$. Lower panel: Same as FIG. 4a, and 4b but for strange quark in FIG. 4c and 4d respectively. Charge neutrality conditions is not imposed here.}
\end{figure}

In FIG. \ref{fig:condensates_amub}a and \ref{fig:condensates_amub}b, we have plotted, for the up quarks, respectively the  dimensionless scalar condensate $-{\langle\bar u u\rangle}/{\Lambda^3} \equiv {I_s^{u}}/{\Lambda^3}$ and dimensionless pseudoscalar condensate ${\langle \bar u i\gamma^5 u\rangle}/{\Lambda^3} \equiv {I^u_{p}}/{\Lambda^3}$ for  as a function of quark chemical potential $\mu_{\rm Q} \equiv \mu_{\rm B}/3$ for different values of axion parameter ${\langle a\rangle}/{f_a} \equiv \theta$. For $\theta=0$ the scalar condensate shows a first order transition at quark chemical potential $\mu_{\rm Q}=\mu_{\rm Q}^c \simeq 361$ MeV. This critical chemical potential decreases with $\theta$. For $\theta=\pi/2$ critical chemical potential turns out to be $\mu_{\rm Q}^c\simeq 357$ MeV. For $\theta=\pi$, the scalar condensate almost vanishes except for the small contribution arising from the finite value of the current quark masses. In FIG. \ref{fig:condensates_amub}b, we show the behavior of pseudoscalar condensates of light quark as a function quark chemical potential for various values of $\theta$. In general, the behavior of $I_p^u$ is complementary to $I_s^u$ regarding the $\theta$ dependence. In vacuum, for $\theta = 0$, $I_p^u = 0$ while for non-vanishing $\theta$, it remains non-zero and becomes maximum at $\theta = \pi$, where the scalar condensate for up quark $I_s^u$ is vanishingly small. For non-vanishing values of $\theta$, as chemical potential increases, the $I_p^u$ also shows a discontinuous behavior similar to $I_s^u$. Further, this transition takes place at the same critical chemical potentials where the chiral transition also takes place. For $\theta = \pi$, the pseudoscalar condensate vanishes at $\mu_{\rm Q} = 345\ {\rm MeV}$. 
This may be contrasted with the behavior of scalar condensates at e.g. $\theta=0$ where $I_s^u$ changes discontinuously to a {\em finite} value for the same due to finite current quark mass. On the other hand,  for $\theta={\pi}/{2}$, the \ac{cp} transition occurs at a higher chemical potential i.e.  $\mu_{\rm Q}^c=357$ MeV where the pseudoscalar condensate shows a discontinuous transition but to a {\em non-vanishing} value for the same similar to the behavior of the scalar condensates $I_s^u$ for all the three values of $\theta$.

We have also plotted the strange quark scalar and pseudoscalar condensates in FIG. \ref{fig:condensates_amub}c and FIG. \ref{fig:condensates_amub}d respectively. The first order chiral transition for light quarks get reflected in a first order transition for the strange quark at the same quark chemical potentials as may be seen in FIG. \ref{fig:condensates_amub}c. This is due to the flavor mixing determinant interaction. However as the chemical potential is increased further, there is a crossover transition for the scalar condensate of strange quarks at $\mu_{\rm Q}^c \sim 480$ MeV. Another interesting point to note here is  that the magnitude of the scalar strange condensate is similar to that of the light quark (up and down) scalar condensates although the current quark masses are very different.  However, the parity violating pseudoscalar condensate for strange quarks is about an order of magnitude less than that of the light quark pseudoscalar condensates. Such a large flavor violation for pseudoscalar condensates could be related to the large mass of the strange quarks as compared to the light quarks which is not reflected in the behavior of scalar condensates. Thus while scalar condensates are not sensitive to the current quark masses, the pseudo-scalar condensates turn out to be rather sensitive to the current quark masses. 

\begin{figure}
    \begin{subfigure}[t]{0.49\textwidth}
      \includegraphics[width=\textwidth]{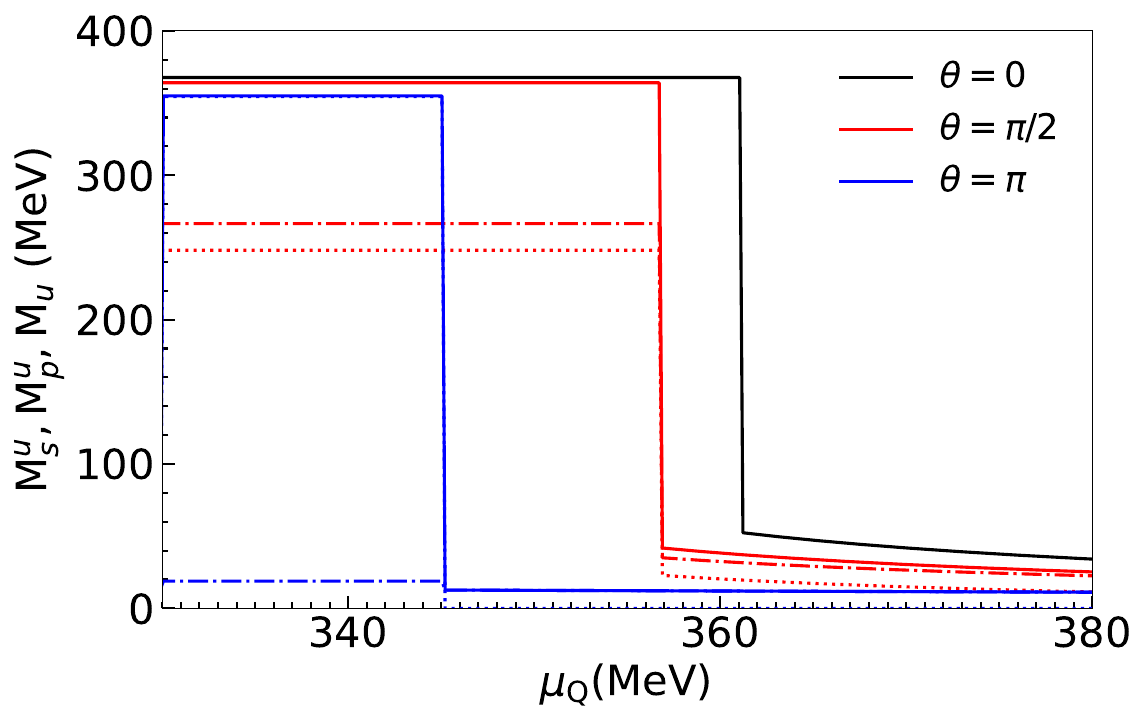}
      \caption{}
    \end{subfigure}
    \begin{subfigure}[t]{0.49\textwidth}
      \includegraphics[width=\textwidth]{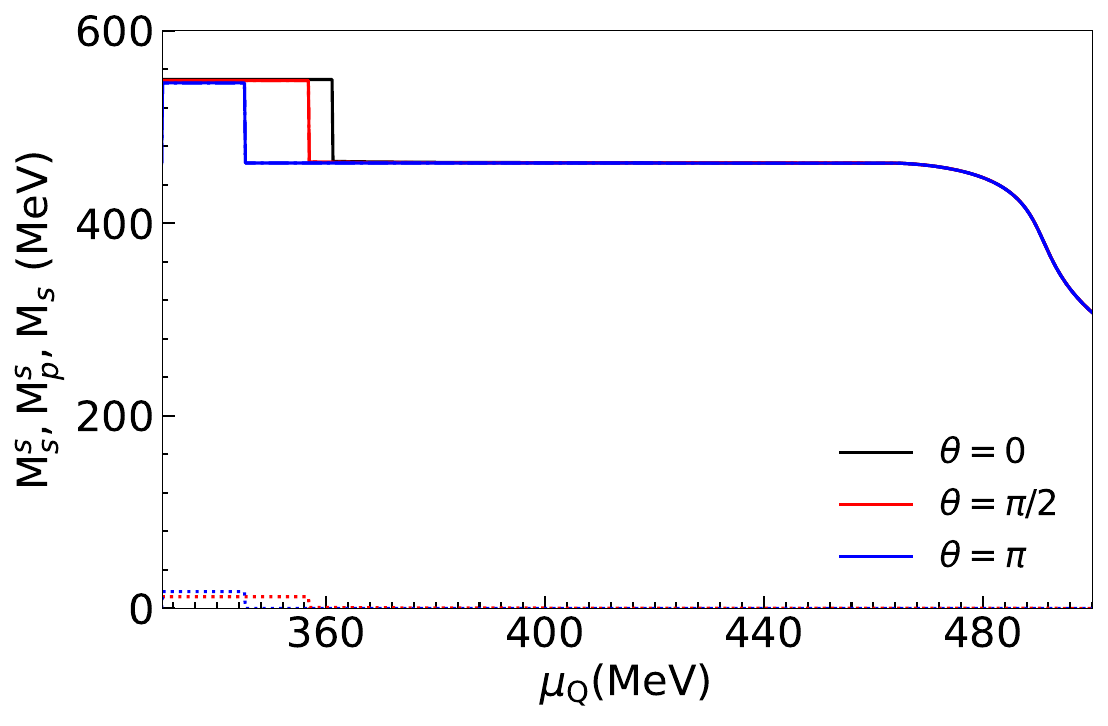}
      \caption{}
    \end{subfigure}
    \caption{\label{fig:quark_masses_amuq} : Different contribution to the constituent quark masses as a function of quark chemical potential $\mu_Q$ for various values of $\theta
    \equiv\langle{a\rangle}/f_a $.$M_s^u$, $M_p^u$ and $M^u=\sqrt {M_s^{u}{}2+M_p^u{}^2}$, the constituent quark masses for up quark arising from scalar condensate, pseudoscalar condensate and the total constituent quark masses respectively are plotted as a function of $\mu_Q$(Fig. 5a). The same quantities for strange quark are plotted in Fig 5}
\end{figure}

 Corresponding to these values of $\theta$ and $\mu_{\rm Q}$, the solutions of mass gap equations, Eq. (\ref{msu}) and Eq. (\ref{mpu}) for $M_s^u$ and $M_p^u$, and the total mass $M^u = \sqrt{M_s^u{}^2+M_p^u{}^2}$ for the up quark is shown in FIG. \ref{fig:quark_masses_amuq}(a) while the same for the strange quarks is shown in FIG. \ref{fig:quark_masses_amuq}(b) for three values of $\theta$. The behavior of the condensates as in FIG. \ref{fig:condensates_amub} gets reflected on the behavior of masses of the light quarks and strange quark.

\begin{figure}
    \begin{subfigure}[t]{0.49\textwidth}
      \includegraphics[width=\textwidth]{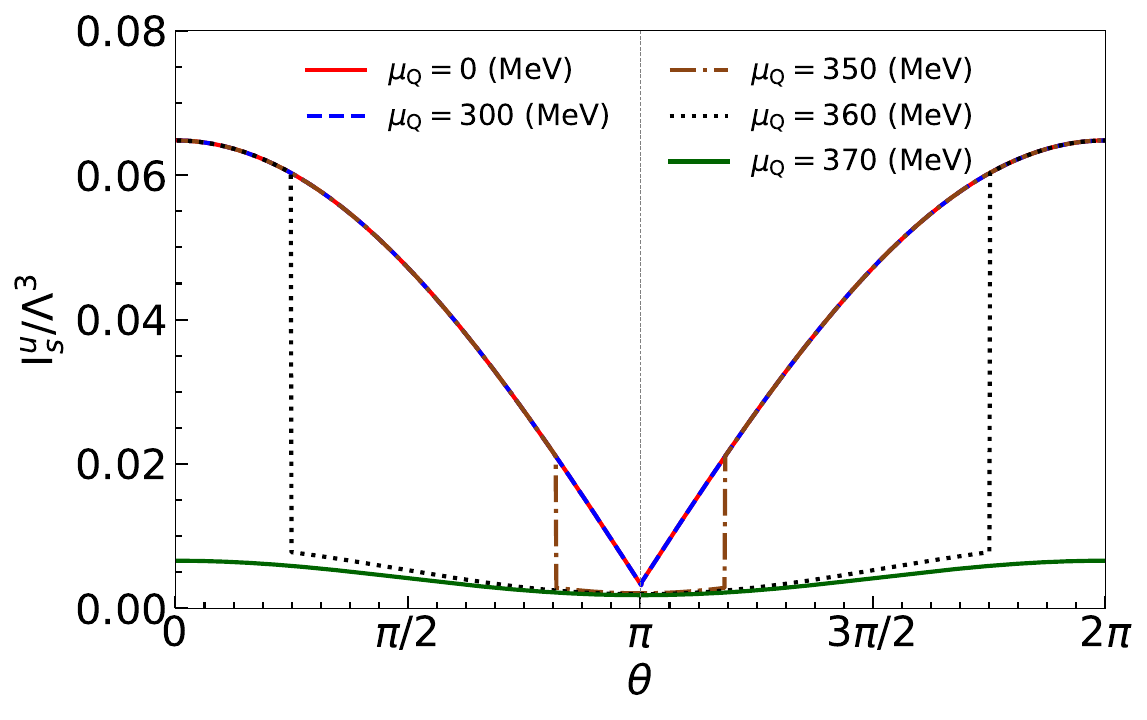}
      \caption{}
    \end{subfigure}
    \begin{subfigure}[t]{0.49\textwidth}
      \includegraphics[width=\textwidth]{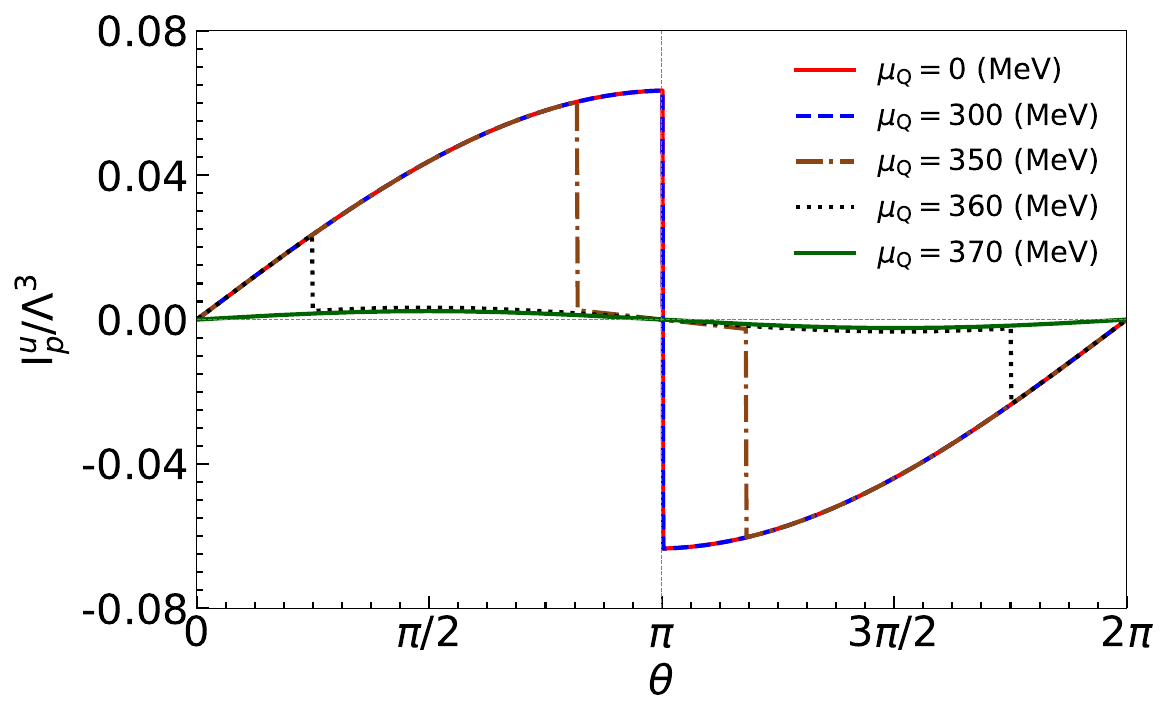}
      \caption{}
    \end{subfigure}
    \begin{subfigure}[t]{0.49\textwidth}
      \includegraphics[width=\textwidth]{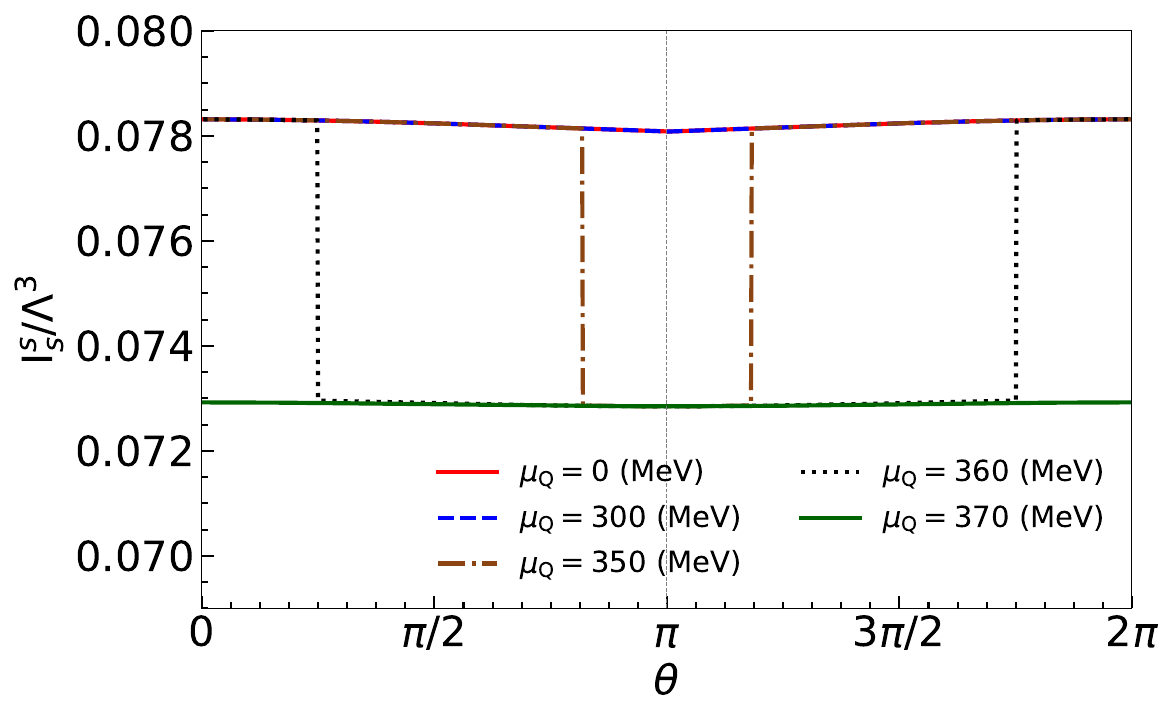}
      \caption{}
    \end{subfigure}
    \begin{subfigure}[t]{0.49\textwidth}
      \includegraphics[width=\textwidth]{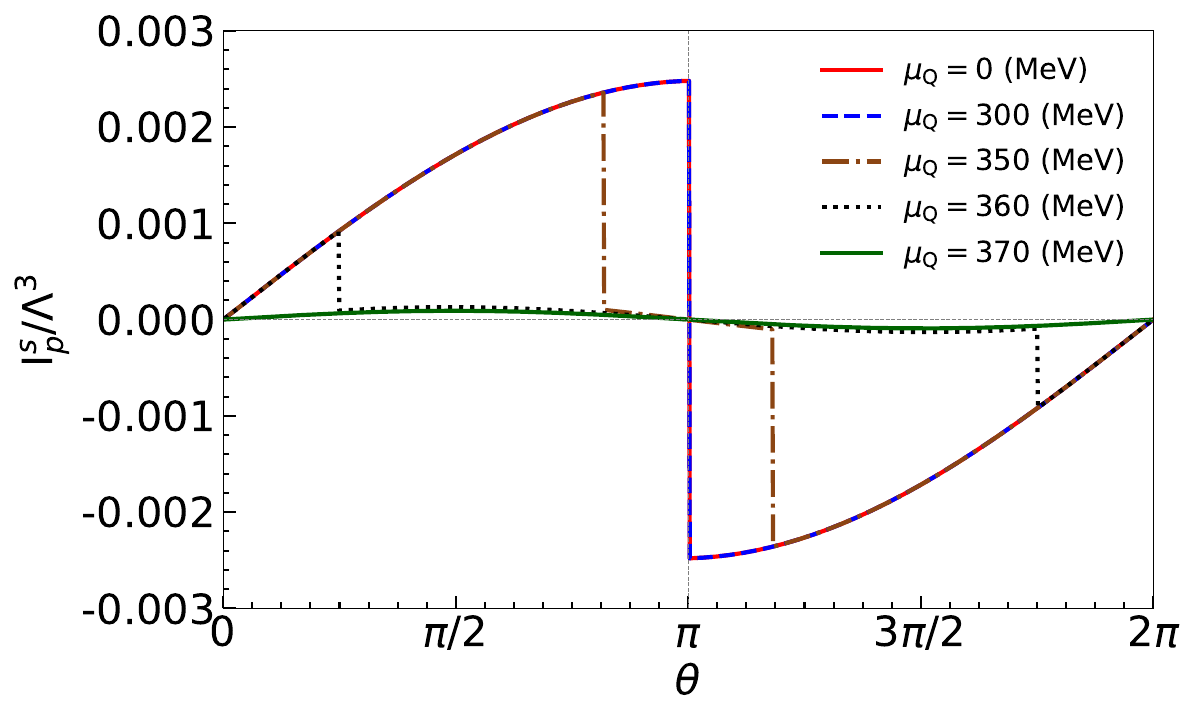}
      \caption{}
    \end{subfigure}
    \caption{\label{fig:condensates_theta} Upper panel: The variations of scalar condensate, $I_u^s/\Lambda^3=\langle{\bar u u \rangle}/\Lambda^3$as a function of scaled axion variable $\theta$  for various values of quark chemical potential (Fig 5a). The discontinuities correspond to chiral restoration.  Fig 5 b: The variation of the pseudoscalar condensate of up quark , $I_p^u/\Lambda^3=\langle{\bar ui\gamma^5 u\rangle}/\Lambda^3$ with $\mu_Q$ .
     Lower panel: the same for heavier strange quark in Fig 5c and Fig 5d respectively. }
\end{figure}
Next, we discuss the behavior of the condensates with the scaled axion field parameter $\theta$.
In FIG. \ref{fig:condensates_theta}(a) we show the variation of $I_s^u$ with $\theta$ for different values for the quark chemical potential i.e. $\mu_Q \in (0,\ 300,\ 350,\ 370)$ MeV. As we mentioned earlier, for $\theta=0$ the critical chemical potential $\mu_{\rm Q}^c=361$ MeV. This chiral restoration chemical potential $\mu_{\rm Q}^c$ decreases for higher values of the axion parameter $\theta$. The plots of $I_s^u$, for $\mu_Q=0$ and $\mu_Q=300$ MeV shown by the red solid and blue solid curves respectively overlap with each other as the chiral symmetry is not restored for any value of $\theta$ in this range of chemical potential. The periodic behavior of this condensate with respect to $\theta$ is due to $\cos \theta$ and $\sin \theta$ dependent terms present in the thermodynamic potential. As $\theta$ is varied, the scalar condensate $I_s^u$ reaches its maximum value for $\theta=2 n\pi$, for $n=0,1,2, \cdots$ and attains a minimum for $\theta=(2n+1)\pi$. As $\mu_{\rm Q}$ increases, e.g. for $\mu_Q=350$ MeV, denoted by the brown dot-dashed curve, the corresponding curve for $I_s^u$ lies on the same curve as for $\mu_{\rm Q}=0$ until $\theta=2.57$ at which point there is a first order transition where the $-\langle \bar u u\rangle$ drops from a value of $(166.5\ {\rm MeV})^3$ to $(85.3\ {\rm MeV})^3$. Beyond this value of $\theta$, the scalar condensate decreases further continuously to its minimum value of $(76.9\ {\rm MeV})^3$ at $\theta=\pi$. Beyond this, the condensate increases with $\theta$ and again makes a discontinuous transition at $\theta=3.71$ to its $\mu_Q=0$ value and follows the same curve for $\mu_Q=0$. The range of $\theta$ value where this continuous behavior for the condensate is seen {\em increases} with increase in $\mu_Q$ . Thus,for $\mu_Q=360 MeV$ the same curve is shown as the dotted black line and its behavior with varying $\theta$ is similar to $\mu_Q=350$ MeV curve except that the {\em critical} value of $\theta\equiv \theta_c=0.78$.  Let us note that for $\theta=0$, the critical chemical potential for chiral restoration is $\mu_Q=361$ MeV. Beyond this value, the condensate does not show any discontinuous behavior but a continuous oscillatory  behavior of as for the case with $\mu_Q=0$, except that the magnitude of the condensate is much smaller. This is essentially seen for the $\mu_Q=370$ MeV denoted by green solid curve in FIG. \ref{fig:condensates_theta}(a). This discontinuous behavior of the scalar condensate with $\theta$ below the chiral transition at $\theta=0$ may be contrasted with the finite temperature and vanishing chemical potential considered earlier \cite{Chatterjee:2011yz,Chatterjee:2014csa, Lu:2018ukl, Bandyopadhyay:2019pml, Das:2020pjg, Gong:2024cwc}.

In FIG. \ref{fig:condensates_theta}(b), we have shown the behavior of the pseudoscalar condensate for up quark, $I_p^u\equiv \langle\bar u\gamma^5 u\rangle$ as a function of axion parameter $\langle a\rangle/f_a\equiv\theta$ for different chemical potential as in FIG. \ref{fig:condensates_theta}(a) i.e. two values below the chiral restoration chemical potential ($\mu_Q=0,\ 300$ MeV) and two values above $\mu_{\rm B}^c$ ($\mu_Q=351,\ 370$ MeV). As may be observed, the pesudoscalar condensate behaves in a complementary manner as compared to the scalar condensate. Contrary to scalar condensates, for $\mu_Q=0$ and $\mu_Q=300$ MeV, the condensate vanishes at $\theta= 2 n\pi$. For $\mu_{\rm Q}=0,\ 300$ MeV, as $\theta$ increases from zero, the magnitude of $I_p^u$ monotonically increases upto $\theta=\pi$ at which point it changes discontinuously to a value equal in magnitude but opposite in sign and then decreases monotonically in magnitude and vanishes at $\theta=2\pi$. At $\theta = \pi$ there are two degenerate vacuum which is in agreement of Dashen \cite{Dashen:1970et} phenomena. The two vacua, which have opposite signs for the condensate $I_p^u$, differ by a \ac{cp} transformation between them. As we shall discuss later, the effective potential as a function of $\theta$ is maximum at $\theta = \pi$. As $\mu_{\rm Q}$ is increased to $350$ MeV, $I_p^u$ changes discontinuously at $\theta = 2.57$ and varies continuously upto $\theta = 3.71$. There is no such discontinuity at $\theta = \pi$. However, below $\theta = 2.57$ and above $\theta = 3.71$ there is a degeneracy in the effective potential with the pseudoscalar condensate $I_p^u$ being equal in magnitude and differing by a sign. Beyond $\mu_{\rm B}^c = 361\ {\rm MeV}$ the condensate $I_p^u$ does not have any discontinuity for any value of $\theta$. This is shown for $\mu_{\rm Q} = 370\ {\rm MeV}$ in FIG. \ref{fig:condensates_theta}(b) by blue solid curve. Beyond $\mu_{\rm B}^c = 361\ {\rm MeV}$, $I_p^u$ becomes a continuous oscillatory function of  $\theta$ and its magnitude decreases with increase in quark chemical potential for all values of $\theta$. 

In FIG. \ref{fig:condensates_theta}(c) and \ref{fig:condensates_theta}(d), we show the same curves for the heavier strange quark. The general behavior here is similar to those for the light quarks except that the variations are much milder which could be a reflection of larger value for the current quark mass of strange quark as compared to the light quarks.

\begin{figure}
    \begin{subfigure}[t]{0.60\textwidth}
    \includegraphics[width=\textwidth]{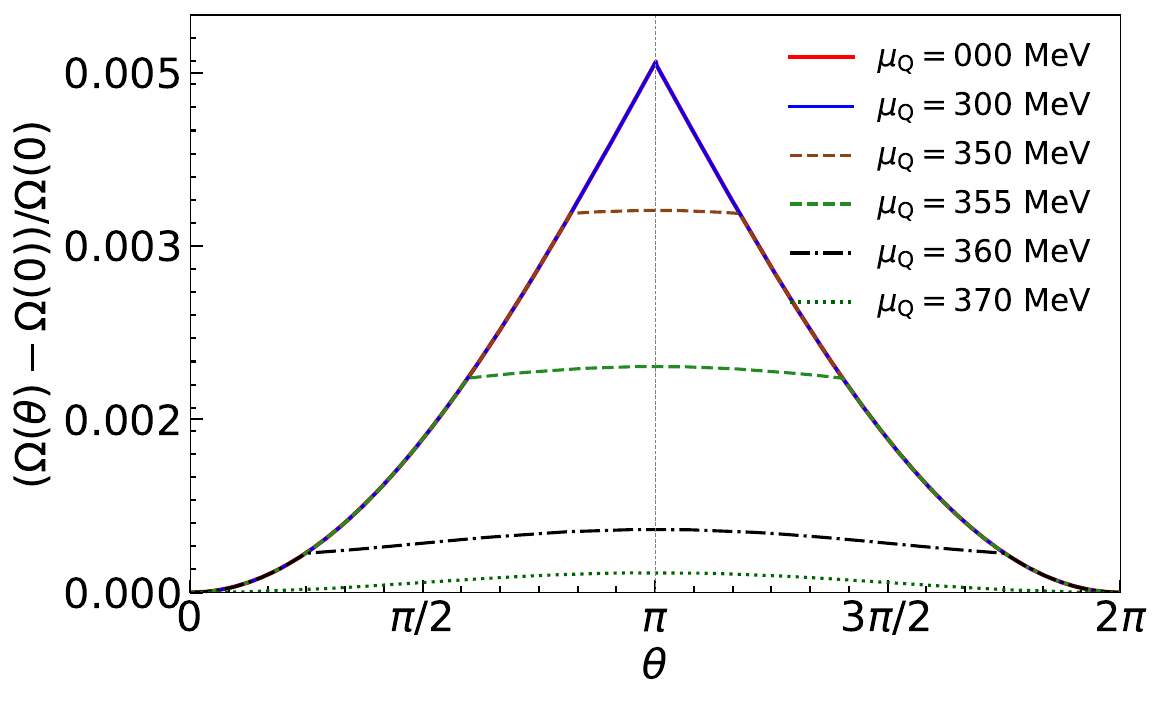}
    \end{subfigure}
\caption{\label{fig:njl3f_omega} Axion potential at finite density. For various quark chemical potentials, the dashed line corresponds to chiral symmetry restored phase and solid line corresponds to chiral symmetry broken phase.}
\end{figure}

In FIG. \ref{fig:njl3f_omega}, we show the variation of the normalized thermodynamic potential or the axion potential as a 
function of $\theta$ for different values of $\mu_{\rm Q}$. For each $\mu_{\rm Q}$, the value of thermodynamic potential at $\theta = 0$ 
has been subtracted. The potential has degenerate vacua at $\theta = 2n\pi$, $n = (0,1,\ 2,\ 3,\ \dots)$. It has also  maxima at 
$\theta = (2n + 1)\pi$, $n = (0,1,\ 2,\ 3,\ \dots)$. As the chemical potential is increased from  r $\mu_Q = 0{\rm MeV}$ to 300 MeV,
 which is below the chiral transition, the potential does not change 
as condensates do not change as shown in FIG. \ref{fig:condensates_theta} . As $\mu_Q$ increased further the height
of the barrier between degenerate vacua 
decreases as chiral symmetry gets restored for some value of $\theta$. This is seen in FIG. \ref{fig:njl3f_omega} for different values of 
$\mu_Q > 350\ {\rm MeV}$. For $\mu_Q > 361\ {\rm MeV}$, the barrier height between degenerate vacua at $\theta = 2n\pi$ 
becomes negligibly small compared to vanishing chemical potential case. 
One may further note that at $\theta = (2n+1)\pi$, although the pseudoscalar condensate vanishes above
 $\mu_Q = 361\ {\rm MeV}$, the thermodynamic potential is still a maximum at $\theta = (2n+1)\pi$. 

In the above discussion for the effect of axions on quark matter regarding chiral symmetry breaking, we have not imposed the 
effects of charge neutrality and of vector interaction. The effect of vector interaction lies in reducing the effective quark 
chemical potentials and therefore the critical chemical potentials for chiral symmetry restoration become larger. 
The qualitative features of the symmetry breaking as a function of axion parameter $\theta$ remains unaffected. 
Imposing charge neutrality on the other hand, removes the degeneracy of chemical potentials of up and down quarks. 
Therefore for the same baryon chemical potential ($\mu_{\rm B} = 3\mu_{\rm Q}$), the chiral symmetry transition for 
down quark takes place earlier as compared to up quarks. Next we shall discuss the consequences of these two effects on 
the equation of state of dense quark matter.

\subsection{\ac{eos} for charge neutral matter and speed of sound with hadron-qaurk phase transition} \label{sub:b}
\begin{figure}
    \begin{subfigure}[t]{0.49\textwidth}
      \includegraphics[width=\textwidth]{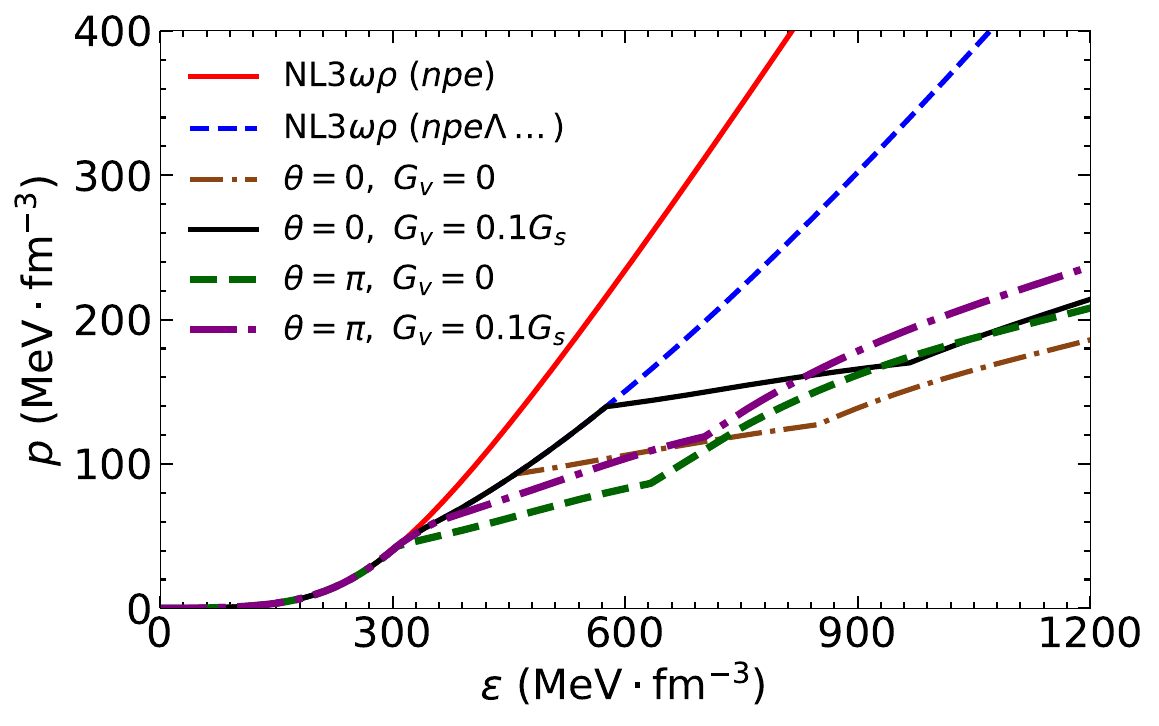}
      \caption{}
    \end{subfigure}
    \begin{subfigure}[t]{0.49\textwidth}
      \includegraphics[width=0.94\textwidth]{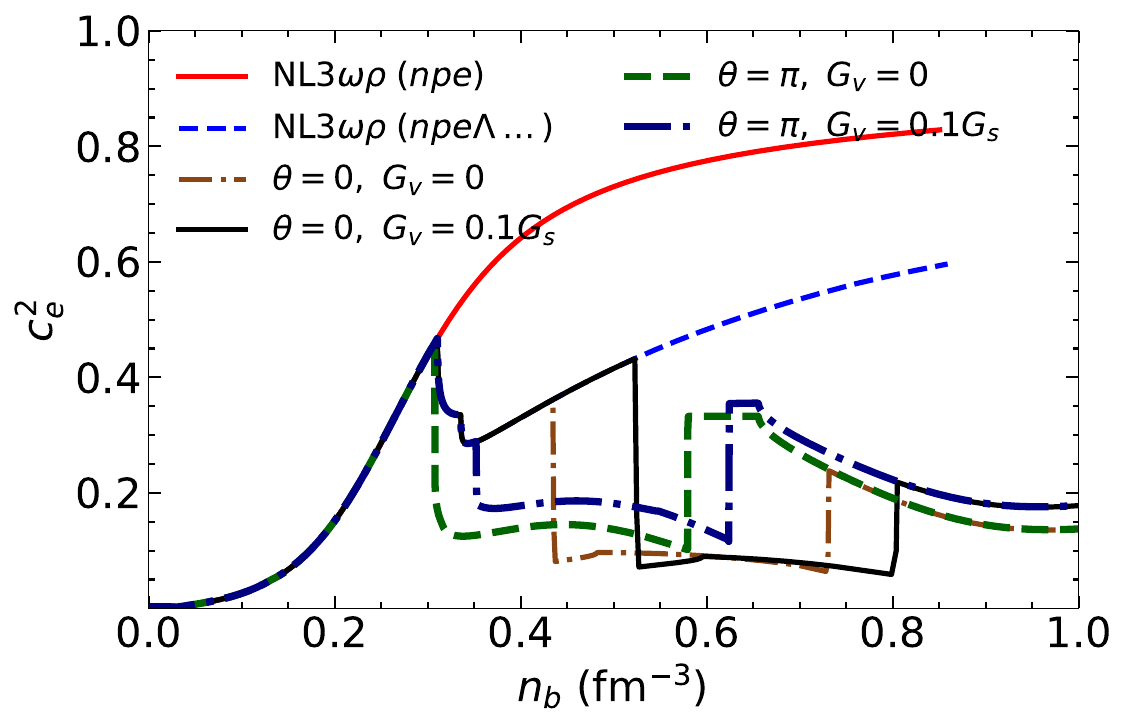}
      \caption{}
    \end{subfigure}
\caption{\label{fig:eos_and_cs2} Fig. 8a :The variation of pressure as a function of energy density with and without phase transitions i.e. \ac{eos} for various values of $\theta$ and $G_{\rm v}$ for the case of quark matter. Fig 8b: The variation of the square of sound speed as a function of baryon number density in the matter with and without phase transition.}
\end{figure}
In FIG. \ref{fig:eos_and_cs2}(a) we display the \ac{eos} with a Gibbs construct for \ac{hqpt} with the three flavors  \ac{njl} \ac{eos} with axions describing the QP and the \ac{rmf} model with NL3$\omega\rho$ parametrization as detailed in TABLE \ref{table.nl3wr.parameters}describing the hadronic phase. The pure hadronic matter \ac{eos}s without hyperons (red solid curve) and with hyperons (blue dashed curve) are  also shown in FIG. \ref{fig:eos_and_cs2}(a) for reference. As expected, the inclusion of hyperons softens the hadronic \ac{eos} as is clearly seen in the figure. For $\theta = 0,\ G_{v} = 0.0$ the \ac{eos} is shown by brown dot-dashed curve. The mixed phase here starts at energy density $\epsilon_1 = 463.7\ {\rm MeV/fm}^{3}$ and ends at $\epsilon_2 = 842.7\ {\rm MeV/fm}^{3}$. The corresponding baryon density for the beginning of the mixed phase $n_{\rm b}^{(1)} \sim 0.44\ {\rm fm}^{-3} \equiv 2.75n_0$, ($n_0 = 0.16\ {\rm fm}^{-3}$) and ends at $n_{\rm b}^{(2)} \sim 0.73\ {\rm fm}^{-3} \equiv 4.56 n_0$ beyond which the matter is in a pure quark matter phase. As we mentioned earlier, the effect of \ac{cp} violation is maximum at $\theta = \pi$. For this value of $\theta$ at $G_{v} = 0$ the corresponding \ac{eos} is shown by the dark-green dashed curve. The threshold for the mixed phase $n_{\rm b}^{(1)}(\theta = \pi) \sim 0.31\ {\rm fm}^{-3} \equiv 1.9n_0$. Increasing the value of $G_v = 0.1 G_s$ makes the \ac{eos} stiffer resulting the threshold for the appearance of the mixed phase to a higher value. For $G_v = 0.1 G_s$ and $\theta = 0$ the \ac{eos} is shown as black solid curve. Here, the threshold energy density and baryon number density ($\epsilon_1,\ n_{\rm b}^{(1)}$) are 579.8 MeV/fm$^3$ and 0.53 fm$^{-3} \sim 3.3n_0$ and reaches upto $\epsilon_2 = 906.2\ {\rm MeV/fm}^{3}$ and $n_{\rm b}^{(2)} = 0.76\ {\rm fm}^{-3} \sim 4.75n_0$. The same for $\theta = \pi$, and $G_v = 0.1 G_s$, the \ac{eos} is shown by the purple dot-dashed curve in the figure. The threshold energy density and baryon density for the appearance of the mixed phase respectively are  $358.7$ MeV/fm$^{3}$ and $0.35\ {\rm fm}^{-3} \sim 2.2n_0$. The mixed phase existed upto $\epsilon_2 = 699.8\ {\rm MeV/fm}^{3}$ and $n_{\rm b}^{(2)} = 0.62\ {\rm fm}^{-3} \sim 3.9n_0$ beyond which the matter is in pure quark matter phase.

In FIG. \ref{fig:eos_and_cs2}(b), we show the variation of square of speed of sound ($c_e^2$) for the charged neutral matter as a function of baryon number density. We have shown the $c_e^2$ for nucleonic, hyperonic, as well as for the mixed phase with different values of scaled axion field $\theta$ and vector repulsion for the quark matter. As the density is increases in the HP the $c_e^2$ increases monotonically for the charge neutral nucleonic matter saturating at about $c_s^2 \sim 0.8$ at high density as shown by the red solid curve here For the hyperonic matter, shown as blue dashed line, the maximum value arises to about 0.5 where the $\Lambda$ hyperons start to appear. With appearance of hyperons the \ac{eos} becomes softer and hence the square of speed of sound shows a drop. Different successive drops in this curve correspond to the appearance of different hyperon species $\Xi^{-}$ and $\Sigma^{+}$. Beyond appearance of $\Sigma^{+}$ hyperon, the square of speed of sound in hyperonic matter increases monotonically as shown in the figure as no other hyperons appear in the density region considered here. Next, we discuss about the sound speed in the mixed phase when there is a \ac{hqpt}.The green dashed curve corresponds to $\theta = \pi$ and $G_v = 0$ case. Here the mixed phase start appearing at $n_{\rm B} = 0.31\ {\rm fm}^{-3}$ which is a little below the $\Lambda$ hyperon threshold for the hypernic matter. One only has the mixed phase of nucleonic matter and quark matter. The sound velocity decreases discontinuously from about $c_e^2 = 0.6$ to $c_e^2 = 0.12$ beyond which it shows a continuous behavior till the end of mixed phase where it again discontinuously increases from $c_e^2 = 0.1$ to  $c_e^2 = 0.33$ of pure quark matter phase. As the density is increases further in the quark matter phase at $n_{\rm B} = 0.67\ {\rm fm}^{-3}$ the sound velocity start decreasing due to appearance of strange quark which softens the \ac{eos} as seen in FIG. \ref{fig:uds_fraction}. For $\theta = 0$ and $G_v = 0$ i.e. without axion field shown by the green dotted curve in the figure. The chiral transition takes place at a higher value of $n_{\rm B}^c = 0.44\ {\rm fm}^{-3}$ and the hyperons $\Lambda$, $\Xi^{-}$ and $\Sigma^{+}$ appear in a hadronic phase. At the on set of the mixed phase for the \ac{hqpt} the square of sound speed drops from $c_s^2 = 0.36$ to $c_e^2 = 0.11$. Beyond which it varies continuously till the end of the mixed phase to $n_{\rm B} = 0.73\ {\rm fm}^{-3}$ at which $c_e^2$ rises discontinuously form $c_e^2 = 0.08$ to $c_e^2 = 0.24$. It may be noted that for the same $G_v$ although axion affects the threshold for \ac{hqpt} it does not affect the velocity of sound. Next we show the effects of non-vanishing $G_v$ i.e. $G_v = 0.1G_s$ on the speed of sound for the charge neutral matter. As mentioned earlier $G_v$ stiffens the \ac{eos} of quark matter leading to a larger threshold for the mixed phase. This is seen in FIG. \ref{fig:eos_and_cs2}(b) for both $\theta = 0$ (black-solid curve) and $\theta = \pi$ (purple dot-dashed curve). At $\theta = \pi$ as well as $\theta = 0$ the $c_e^2$ in the quark matter remains the same. Thus finite $G_v$ affects {\em both} the threshold for \ac{hqpt} and the sound speed while inclusion of non-vanishing $\theta$ affects the threshold for the \ac{hqpt} to a lower value of density while it does not affect the sound velocity. It may be noted that such a behavior was also observed for the two flavor \ac{njl} model studied in Ref. \cite{Lopes:2022efy}.

\subsection{Neutron star's mass-radius and $f$-mode oscillations in presence of axions} \label{sub:c}
To solve the \ac{tov} equations one also needs a separate treatment for the \ac{eos} describing the crust matter of \ac{ns}s at very low density. For the outer crust we include a Baym-Pethick-Sutherland (BPS) \ac{eos} \cite{Baym:1971pw}. The outer crust and the outer core are joined together using a polytrope $p(\epsilon) = a + b\epsilon^c$ in order to construct the inner crust, where $c=4/3$ and $a$ and $b$ are determined in such a way that the \ac{eos} for inner crust matches with the outer crust at $n_{\rm B} = 10^{-4}\ {\rm fm}^{-3}$ and the outer core at $n_{\rm B} = 0.04\ {\rm fm}^{-3}$ \cite{Carriere:2002bx}. It is important to note that the differences in \ac{ns}s radii between this treatment of the inner crust EOS and the unified inner crust description including the pasta phases have been found to be less than 0.5 km, as discussed in \cite{Malik:2022zol}.

\begin{figure}
    \begin{subfigure}[t]{0.60\textwidth}
      \includegraphics[width=\textwidth]{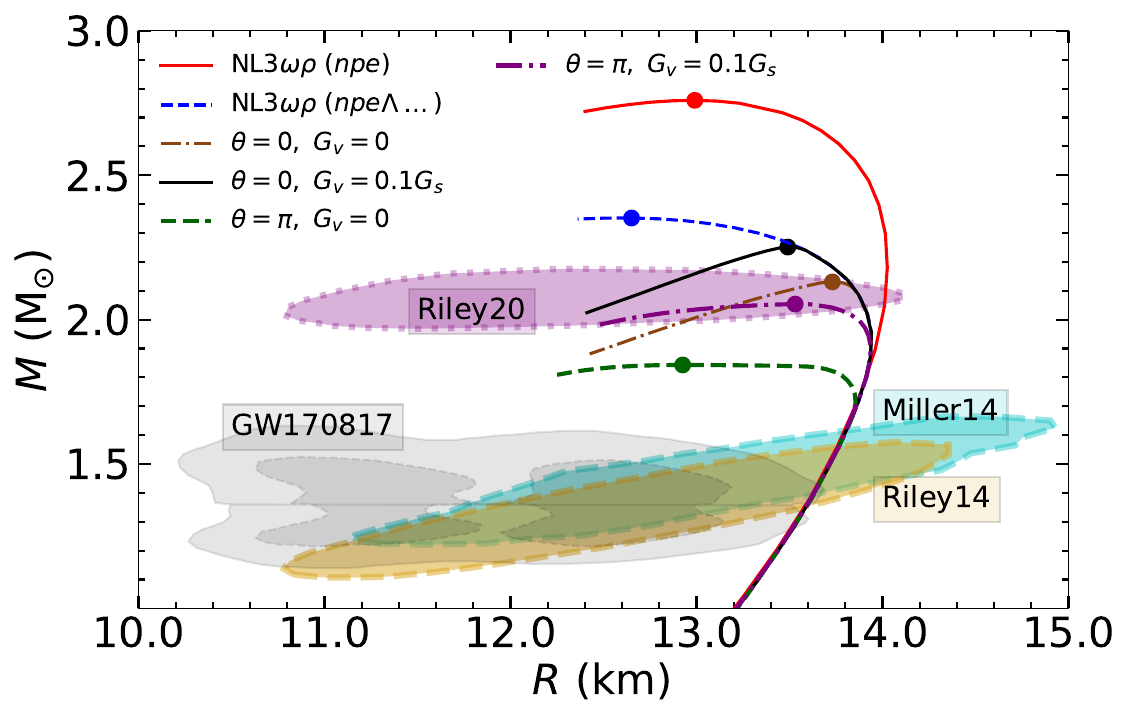}
    \end{subfigure}
    \caption{\label{fig:amr} The mass radius curves for different \ac{eos} depicted in the FIG. \ref{fig:eos_and_cs2} along with the various astrophysical observations. The gray region corresponds to  GW170817 observation \cite{LIGOScientific:2018hze}. The  dark gray and light gray region here correspond to  50\% and 90\% confidence interval (CI)respectively. The magenta patch with dotted outline is the NICER observation of pulsar PSR J0740+6620 of mass $2.14^{+0.1}_{-0.09}\ {\rm M}_{\odot}$ \cite{Riley:2021pdl} while the yellow and blue with dashed outlines are the NICER observations of a pulsar PSR J0030+0451 \cite{Riley:2019yda, Miller:2019cac} of mass about $1.4\ {\rm M}_{\odot}$.}
\end{figure}
An important constraint to be fulfilled by \ac{eos}s is the value of maximum mass which is to be compatible with the observational data. In the FIG. \ref{fig:amr} (a) we present the mass-radius relationships obtained from the \ac{eos}s with/without quark matter core and for hybrid \ac{ns}s with/without axions and different values of $G_v$ for the quark matter \ac{eos}. The largest mass observed upto now, $2.14^{+0.1}_{-0.09}\ {\rm M}_{\odot}$ at 68\% confidence interval for the object PSR J0740+6620 shown as the violet band for the NICER x-rays data \cite{Riley:2021pdl}. For completeness, we also display the Bayesian parameter estimation of the mass and equatorial radius of the millisecond pulsar PSR J0030+0451 as reported by NICER mission. The $M,\ R$ values inferred from the analysis of collected data shown as cyan and yellow zones are $1.36^{+0.15}_{-0.16}\ {\rm M}_{\odot}$ and $12.71_{-1.19}^{+1.14}\ {\rm km}$ \cite{Riley:2019yda},  $1.44^{+0.15}_{-0.14}\ {\rm M}_{\odot}$ and $13.02^{+1.24}_{-1.06}\ {\rm km}$ \cite{Miller:2019cac}. Apart from the NICER data, we also display the constraints from data extracted from LIGO/Virgo gravitational wave observations GW170817. The top and bottom gray regions indicate the 90\% (solid) and 50\% (dashed) confidence intervals of LIGO/Virgo analysis for each binary component of GW170817 event \cite{LIGOScientific:2018hze}. We have considered the cases of \ac{eos}s for (1) $npe\mu$ matter, (2) hyperonic matter, (3) hyperonic matter with quark matter. Further, for quark matter, in particular, we analyze the cases $\theta = 0$ and $\theta = \pi$ so that the results can be shown in the absence of axion effect and when these play large role. We also take two values of $G_v$ i.e. $G_v = 0$, and $G_v = 0.1G_s$ to study the interplay of axions and the vector interaction in the \ac{njl} model. As may be seen in the FIG. \ref{fig:eos_and_cs2}(a) all the six cases we have considered satisfy the constraints from astrophysical observations except the case of hybrid \ac{ns}s with $\theta = \pi$ and $G_v = 0$. It is observed that with this parametrization of hadronic \ac{eos} a pure hadronic star with/without hyperons is consistent with both the NICER and GW observations. It may be noted that without $\omega\rho$ crossed coupling interaction simple NL3 parametrization does not satisfy the GW170817 constraint \cite{Kumar:2021hzo}. The maximum mass for $npe$ matter turns out to be $2.76\ {\rm M}_{\odot}$ while the same for hyperonic matter $2.35\ {\rm M}_{\odot}$ as represented by filled circles (orange, and red respectively) in the figure. Next, we discuss the hybrid \ac{ns}s for $G_v = 0$ and $\theta = 0$ shown by the purple dot-dashed curve where we have hybrid \ac{ns}s with hadronic matter and quark matter core. Because of softer quark matter \ac{eos} the maximum mass here is about $2.13\ {\rm M}_{\odot}$. When the vector interaction is non-zero i.e. $G_v = 0.1G_s$ the corresponding curve is shown by the solid-brown curve and maximum mass of \ac{ns} increases to $2.25\ {\rm M}_{\odot}$. Next we come to the case where the axion effects play a larger role i.e. $\theta = \pi$ case. Here, for $G_v = 0$ is shown by dashed-pink curve and the maximum mass becomes $1.84\ {\rm M}_{\odot}$ and does not satisfy the constraint from the NICER observation. On the other hand as $G_v$ increases to $G_v = 0.1G_s$ (shown by the dotdashed-purple curve) the maximum mass of such a hybrid \ac{ns} becomes $2.05\ {\rm M}_{\odot}$ ans satisfies the maximum mass constraint. Thus a vector interaction for the quark matter becomes essential to have a stable hybrid \ac{ns} with hyperonic outer-core and axionic quark matter as a inner core.

\begin{figure}
    \begin{subfigure}[t]{0.49\textwidth}
      \includegraphics[width=0.94\textwidth]{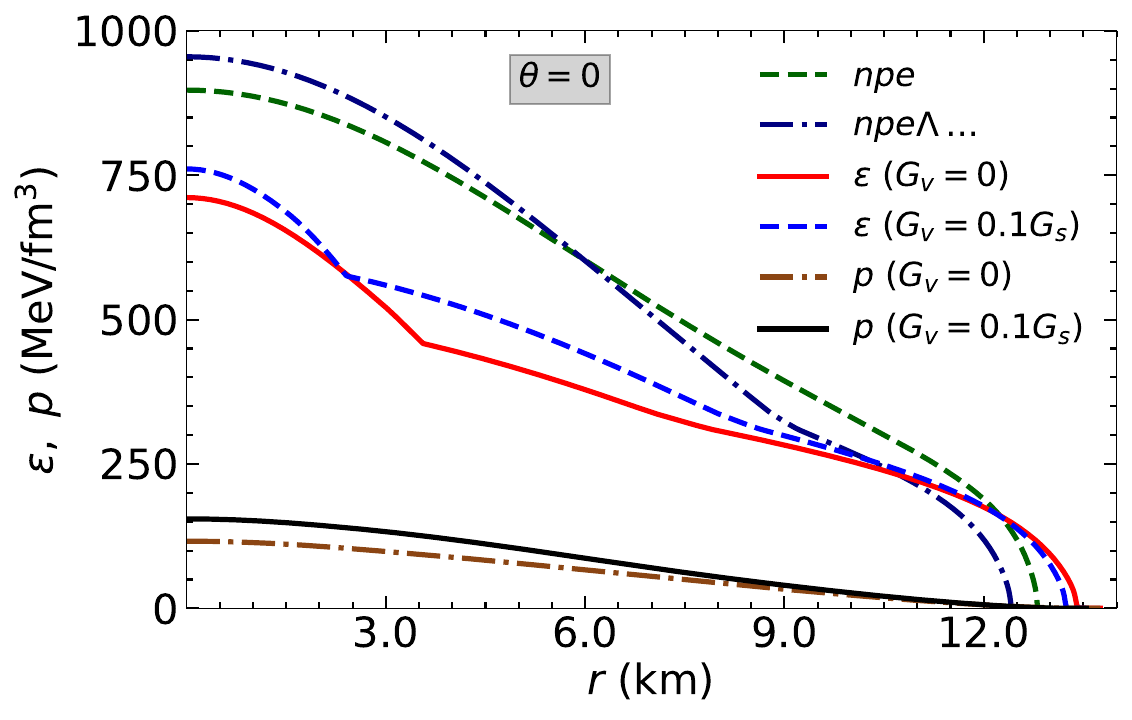}
      \caption{}
    \end{subfigure}
    \begin{subfigure}[t]{0.49\textwidth}
      \includegraphics[width=0.94\textwidth]{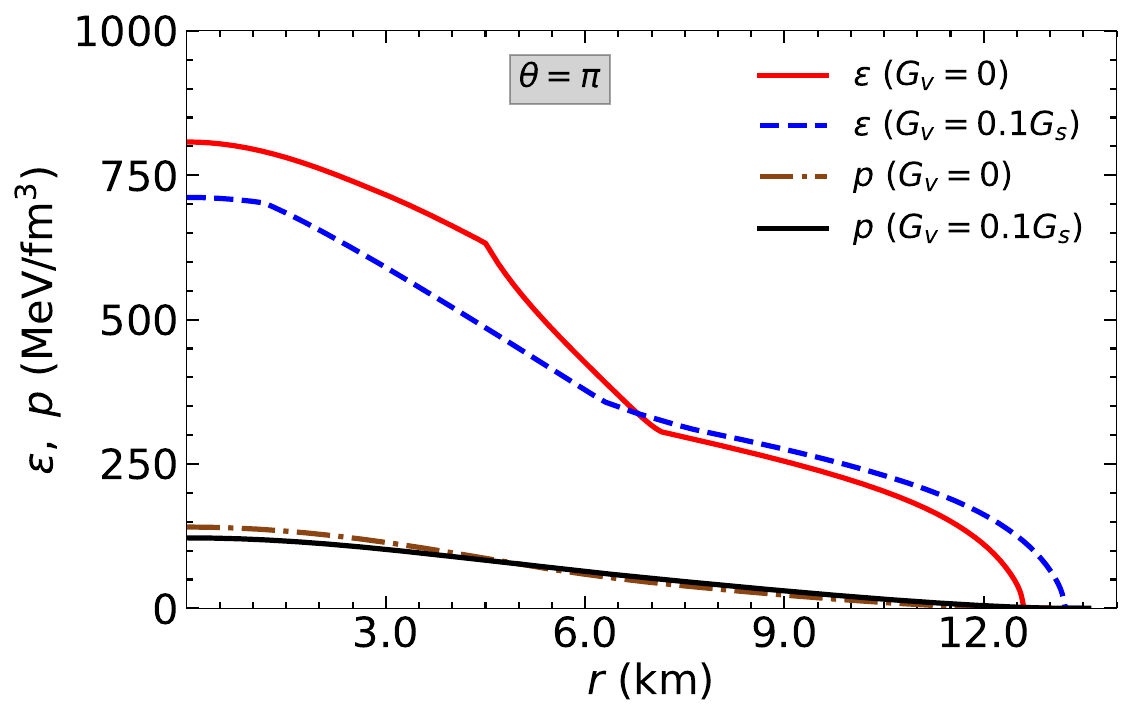}
      \caption{}
    \end{subfigure}
\caption{\label{fig:profile_eps_and_psr} The profiles of the energy density $\epsilon(r)$ and the pressure $p(r)$ for compact stars corresponding to maximum mass as a function of radial distance from the center. The curves with kinks in the energy density correspond to hybrid stars.  The  position of the kinks in the energy density curves correspond to the end of quark matter in mixed  phase. Fig. 10a shows the profiles without axions ($\theta=0$)
while Fig. 10b corresponds to $\theta=\pi$. The two kinks for the case of $\theta=\pi$
and $G_v=0.1 G_s$ correspond to the start and end of the mixed phase of quark matter.For comparison we have also shown in Fig.(a), the energy density and pressure profiles for nucleonic star (purple dot-dashed curve) and hyperonic stars (green dashed curve).}
\end{figure}
Next, we show the energy density and pressure profiles as a function of radial distance from the center of hybrid \ac{ns}s of corresponding maximum mass in FIG. \ref{fig:profile_eps_and_psr}. In FIG. \ref{fig:profile_eps_and_psr}(a), we also display the energy density profiles for a nucleonic as well as a hyperonic \ac{ns} for the comparison. Here, we show the case of hybrid \ac{ns}s without axion i.e. $\theta = 0$ for different values of vector coupling $G_v$ for quark matter. We have plotted the profiles for the corresponding stable maximum mass hybrid \ac{ns}s. In case of $\theta = 0$ with both the values $G_v$ the quark matter in the inner core of hybrid \ac{ns}s is in a mixed phase. The kink in the energy density profile indicates the interface where the mixed phase starts as one  comes the surface to the center of the star. For $G_v = 0$, the star has mass of 2.13 M$_{\odot}$ and a radius of 13.7 km with a quark matter core in the mixed phase with a radius of 3.56 km. As $G_v$ is increased the size of quark matter core decreases to a radius of 2.9 km with a total radius 13.5 km. In FIG. \ref{fig:profile_eps_and_psr}(b) we show the same profiles for the case of $\theta = \pi$. For $G_v = 0$, one has a larger core of quark matter core of radius of 7.3 km of which inner core of 4.7 km in the pure quark matter phase. The total mass of such a star is 1.84 M$_{\odot}$ and radius 12.9 km. Let us note that such a hybrid \ac{ns} with pure quark matter core does not satisfy the maximum mass constraint of two solar mass. As we increase $G_v = 0.1 G_s$ for $\theta = \pi$, the maximum mass becomes 2.05 M$_{\odot}$ with a radius 13.5 km satisfying the constraint of two solar mass star. Such a star, however, can have a pure quark matter core of radius 1.6 km. This has also quark matter component upto a radius 6.5 km beyond which the matter constitute charge neutral hyperonic matter. Thus inclusion of axions can make hybrid \ac{ns} with a pure quark matter core gravitationally stable. Such a conclusion is in line with Ref. \cite{Lopes:2022efy} where the \ac{hqpt} was attempted with a Maxwell construct. The Gibbs construct as considered here can have the possibility of hybrid \ac{ns}s with a quark matter core in mixed phase even without axion. 

\begin{figure}
    \begin{subfigure}[t]{0.49\textwidth}
      \includegraphics[width=0.94\textwidth]{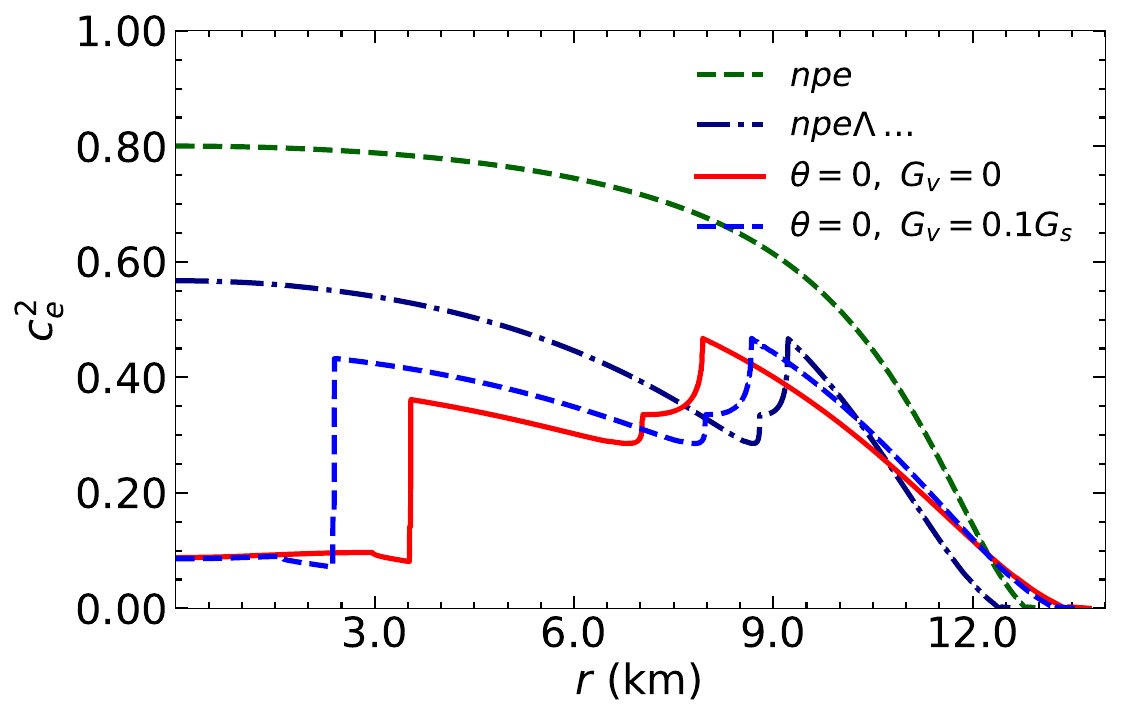}
      \caption{}
    \end{subfigure}
    \begin{subfigure}[t]{0.49\textwidth}
      \includegraphics[width=0.94\textwidth]{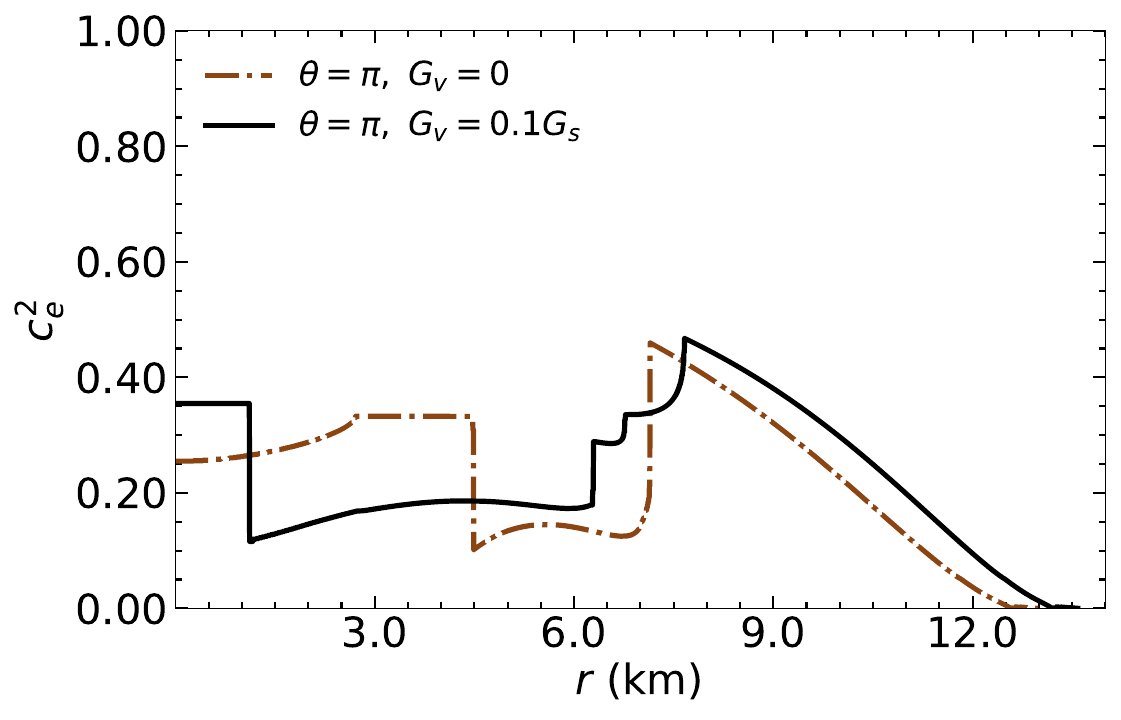}
      \caption{}
    \end{subfigure}
\caption{\label{fig:profile_cs2} The variation of speed of sound inside a compact star as a function of radial distance from the center of a corresponding maximum mass compact\ac{ns}. The left figure (a) corresponds to hybrid stars without axions ($\theta=0$) while the right figure (b) corresponds to hybrid star with axions ($\theta=\pi$). For comparison, nucleonic (green dashed line) and hyperonic (purple dot dashed line) \ac{ns} sound speed profiles are also shown in Fig. 11a. Discontinuous changes of $c_e^2$ for hybrid \ac{ns} represent a \ac{hqpt} in the core. The little jumps in sound speed  in the hyperonic matter phase correspond to the disappearance of different species of hyperons with increasing $r$.}
\end{figure}
Next, we shall discuss the variation of speed of sound as a function of distance from the center of the star. But before that a few remarks regarding the same may be in order. Let us note that constraints on the radius R of \ac{ns} from gravitational wave data suggested that $R < 13.5$ kms which,in turn, means the \ac{eos} is soft at nuclear saturation densities \cite{Annala:2017llu, Tews:2018iwm}. On the other hand, observations of high mass pulsars with masses greater than two $M_{\odot}$ indicate the \ac{eos} to be sufficiently stiff at larger densities \cite{Raaijmakers:2019dks, Drischler:2020fvz}. Such constraints on \ac{eos} also lead to an interesting feature of speed of sound in dense matter. Theoretically, one can estimate the speed of sound at low densities using chiral effective theory ($\chi$EFT) while at asymptotically high densities using perturbative QCD in a controlled manner. This tells us that the speed of sound $c_e^2<<1$ at low densities and it approaches the conformal bound $c_e^2={1}/{3}$ from below at very high densities. At intermediate densities, neither the perturbative QCD nor the $\chi$EFT is applicable and the speed of sound can be monotonic or nonmonotonic function of baryon density. The constraints on \ac{eos} as obtained from \ac{ns} measurements reveals that the $c_e^2$ is, most likely a nonmonotonic  function of density violating the conformal bound at few times the saturation density of nuclear matter. In particular, $c_e^2$ need to increase monotonically from nuclear matter density to few times nuclear matter density overshooting the conformal value reaching a peak at some intermediate density. As the density is increased further, $c_e^2$ should decrease reaching below the conformal bound to a minimum and eventually increasing to reach the conformal limit asymptotically. There have been different approaches e.g. using a hadronic-quarkyonic cross over transition \cite{McLerran:2018hbz, Pang:2023dqj} or with a vector condensate along with pionic fluctuations \cite{Pisarski:2021aoz} to explain the origin for such a behavior of sound speed in dense matter. Similar behavior for the sound speed is obtained here through \ac{hqpt} with a mixed phase construct for the same. 

With these remarks for the speed of sound in dense matter, in FIG. \ref{fig:profile_cs2}(a), we display the speed of sound profile inside the hybrid \ac{ns}s without axions ($\theta=0$) and in Fig \ref{fig:profile_cs2}(b) with axions ($\theta=\pi$) { as a function of radial distance inside the hybrid \ac{ns}}. In FIG. \ref{fig:profile_cs2}(a), we also display the sound speed profiles of hadronic stars for comparison with the nucleonic \ac{ns} shown as the green dashed curve while the hyperonic \ac{ns} being shown as the purple dashed dot lines. All the curves here refer to the corresponding maximum stars. As may be observed for the nucleonic star, the speed of sound shows a monotonic rise from the surface to the center where it becomes about $c_s^2=0.8$ signifying the stiffness of the \ac{eos} to support the maximum mass star with a mass of 2.7M$_{\odot}$. The speed of sound for hyperonic star shown by the purple dot-dashed courve on the other hand reaches the maximum value of $c_e^2=0.58$ at the center as hyperonic matter is softer than that of nucleonic matter. The sudden drops in the speed of sound as one moves from the surface to the center,as displayed here,correspond to appearance of different hyperons. On the other hand, for hybrid stars the drop in sound speed is more dramatic. For $\theta = 0$ and $G_v = 0$, the sound speed profile is shown by the solid- red line. At the center of the star, the matter is in a mixed phase with $c_e^2\simeq 0.1 $ which varies continuously till the mixed phase ends at $R = 3.5\ {\rm km}$ where it jumps discontinuously to a value 0.36. With increasing the radial distance from the center the speed of sound decreases and jumps up again discontinuously twice signifying disappearance of $\Sigma^{+}$, $\Xi^{-}$ and $\Lambda$ hyperons to reach a maximum value of $c_e^2\simeq 0.47$ and finally decreases smoothly with radial distance and vanishes at the surface. Similar behavior is also seen for $G_v = 0.1G_s$ shown by blue-dashed  curve except that the mixed phase quark matter core here is smaller which is due to the fact that the the quark matter \ac{eos} becomes stiffer with $G_v$ and the phase transition  takes place at a larger baryon density compared to the case without the vector interaction. The case of including the axions ($\theta=\pi$) is shown in Fig \ref{fig:profile_cs2}(b). For $\theta = \pi$ and $G_v = 0$, is shown as the brown dot dashed curve. Here, in the quark matter phase the velocity of sound changes discontinuously twice one at the beginning of the mixed phase and the other is at the transition to the pure quark matter core. As one moves towards the center after the onset of pure quark matter phase, the speed of sound remains constant and then starts decreasing near the center which is due to the onset for the appearance of strange quarks in the core. In this case, the hadronic phase does not have hyperons. 

When the vector interaction  is increased to $G_v=0.1G_s$, shown by the black solid line, there are again two discontinuities for speed of sound in  the quark matter core - one at the onset of the mixed phase and the other on the onset of the pure quark matter phase. The pure quark matter phase in this case  has up and down quarks without the strange quarks. In the hadronic matter phase, unlike the case for $G_v=0$, $c_e^2$ jumps up  twice at the onset of disappearance of hyperons $\Xi_-$ and $\Lambda$ as one moves towards the surface  from the center. Let us note that the variation in sound velocity in the interior of the stars plays a crucial role in determining the characteristics of various \ac{nro} modes that we discuss next.

\begin{figure}
    \begin{subfigure}[t]{0.60\textwidth}
      \includegraphics[width=0.94\textwidth]{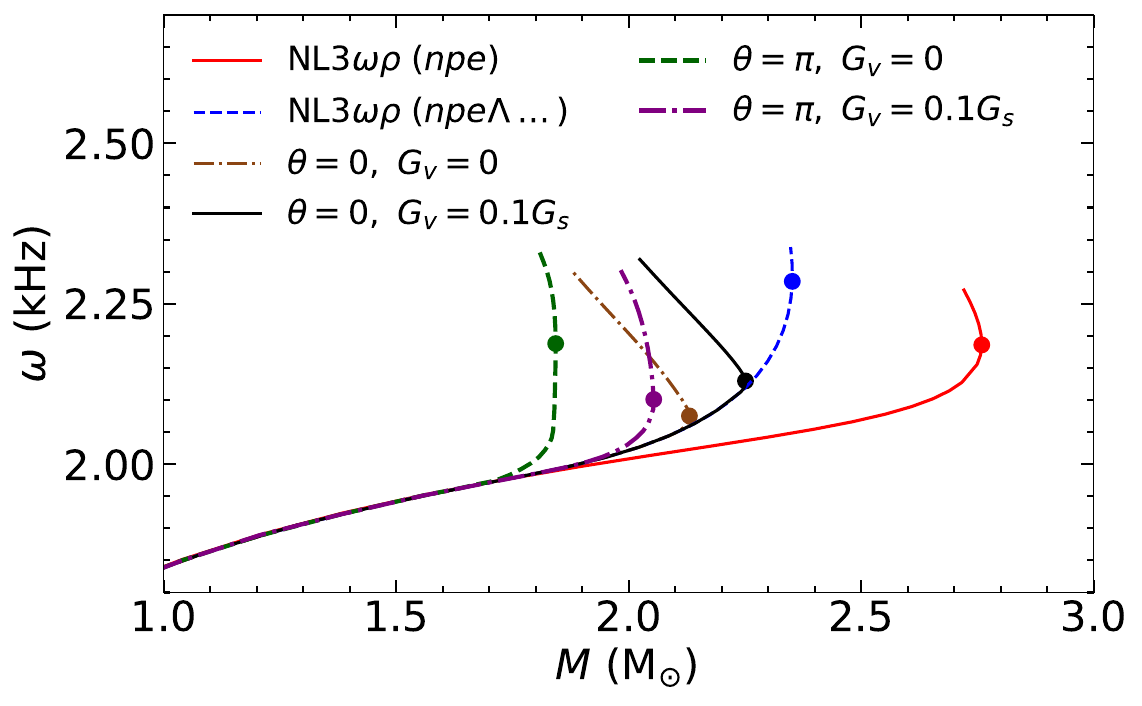}
    \end{subfigure}
    \caption{\label{fig:omega} The frequencies of quadruple fundamental modes  ($f$-mode) of \ac{nro} as a function of \ac{ns} masses. The filled dots correspond to the maximum mass
    of the stable compact stars.  The enhancement at the onset of quark matter core may be noted.}
\end{figure}
In FIG. \ref{fig:omega}, we display the \ac{nro} frequencies for the quadrupole $f$ modes for different compositions of compact stars. The filled circle in each of these curves correspond to the maximum mass with the corresponding composition. In general, the mode frequency increases with stellar mass for all the different composition of \ac{ns}. The solid-red curve corresponds to $npe$ matter. The maximum frequency for the stable stable star with nucleonic \ac{eos} turns out to be 2.18 kHz corresponding to $M_{\rm Max} = 2.76\ {\rm M}_{\odot}$. The case of hyperonic star without any quark core is shown by the dashed-blue curve with a $M_{\rm Max} = 2.35\ {\rm M}_{\odot}$ having the corresponding $f$-mode frequency $\omega = 2.31$ kHz. As may be seen in the Fig. \ref{fig:omega}, for the same mass of  a purely neucleonic star $\omega = 2.05$ kHz. Inclusion of hyperons thus increases the $f$ mode frequency by about 260 Hz. Such an enhancement of $f$ mode frequencies with hyperons was also observed earlier \cite{Pradhan:2022vdf, Pradhan:2020amo}. 
 
Next we discuss the case of  hybrid \ac{ns} with quark matter core. Without axions, i.e. $\theta = 0$ and for vanishing vector interaction $G_{\rm v} = 0$, the corresponding curve is shown by dash-dotted-brown line. In this case, $M_{\rm max}=2.13 {\rm M}_\odot$ and the corresponding $f$-mode frequency is $\omega=2.08$ kHz. A nucleonic star of same mass will have $\omega=2.02$ kHz. Thus, there is a  60 Hz enhancement due to quark matter core in mixed phase with nuclear matter. When the vector coupling is increased to $G_{\rm v}=0.1 G_{\rm s}$, the maximum mass increases to $M_{\rm max}=2.25\ {\rm M}_\odot$ and corresponding $\omega=2.13$ kHz which is a 90 Hz enhancement compared to a nucleonic \ac{ns} of same mass. Let us note that dominant contribution this hike in $\omega$ arises from hyperons as the quark matter contribution is smaller as the quark matter core is smaller.

Next, we discuss the case when the effect of axions is maximal i.e. $\theta=\pi$. For vanishing vector coupling i.e. $G_{\rm v}=0$, such a hybrid star will have a larger quark matter core with the quark matter being in pure phase as well as in mixed phase with hyperonic matter having a maximum mass of 1.84 ${\rm M}_\odot$. In this case, for $M_{\rm max}$, the $f$-mode frequency is $\omega=2.19$ kHz which is a large enhancement of 210 Hz compared to a nucleonic star with the same mass. Such a case is however not satisfy the maximum mass constraint. As the vector coupling is increased to $G_{\rm v}=0.1 G_{\rm s}$, the maximum mass becomes $M_{\rm max}=2.05\ {\rm M}_\odot$ consistent with maximum mass constraint and the corresponding $\omega=2.1$ kHz. This is about 110 Hz enhancement compared to a nucleonic star of the same mass. This large enhancement for $\theta = \pi$ and $G_{\rm v} = 0.1G_{\rm s}$ corresponding to the presence of axions is due to two reasons. First of all for the hadronic phase the enhancement is due to presence of hyperons and there is a further enhancement due to a contribution from the quark matter in a larger core, both in a pure phase and in a mixed phase with hyperonic matter as compared to absence of axions ($\theta = 0$). Presence of axions thus not only stabilizes the hybrid star but also leads to a substantial enhancement in $f$ mode frequency.

It may be relevant here to discuss the corrections to the $f$ mode frequencies due to the Cowling approximation as compared to performing a linearized general relativity estimation including a metric perturbation. In this context, it may be noted that in Ref. \cite{Yoshida:1997bf}, a comparative analysis was performed between $f$ mode frequencies obtained from the linearized GR and the Cowling approximation. Such an analysis showed that for compactness ($C=M/R$) $C=0.05$ the $f$ mode frequency is overestimated by 30\%; but it decreases to 15\% for $C=0.2$ \cite{Yoshida:1997bf}. A recent analysis in Refs \cite{Zhao:2022tcw, Rather:2024nry} gives similar range with the error decreasing with increasing compactness. For the hybrid stars with axions as considered here the compactness parameter is little larger than 0.2. Therefore, we expect the errors in estimation of $f$ mode frequencies within the Cowling approximation for the hybrid stars are no more than 15-20\%.

\section{Summary and conclusion \label{summary_and_conclusion}}
The objective of the present investigation has been two fold. The first is to study the effects of \ac{cp} violation with axions on \ac{qcd} phase diagram at zero temperature and finite density. The second objective is to investigate the effects of axionic quark matter on \ac{hqpt}, \ac{ns} structure and their non-radial oscillations. To that end, we calculated the effects of axions on the chiral transition in dense matter. The interaction of axions with \ac{qcd} matter is modeled by a three flavor local \ac{njl} model. This has been considered earlier within a two flavor \ac{njl} model but near the chiral transition at finite temperature and finite baryon density in the context of axion domain walls \cite{Zhang:2023lij}. For the three flavors case, it turns out that the parity violating strange quarks condensate is about an order of magnitude smaller than that for the light quarks. Nonetheless, the strange quark condensate in the scalar channel affects the light quarks condensates in the pseudoscalar channel in a significant manner through a flavor mixing determinant interaction. 

The presence of axions reduces the critical density for the chiral transition. The effect of first order chiral transition is seen in the axion potential at zero temperature and finite density. The vacuum  effective potential for axions with a sharp peak at $\theta=\pi$, becomes flatter in the presence of matter around $\theta=\pi$ for the values of $\theta$ where chiral crossover transition takes place. This results in a reduction of the height of the potential between degenerate vacuum $\theta=2n\pi$. We also studied the behavior of the scalar and pseudoscalar condensates as a function of axion variable $\theta$. In vacuum, the pseudoscalar condensate shows a discontinuous behavior at $\theta=\pi$ consistent with Dashen phenomenon. As chemical potential is increased, this discontinuity begin to occur at $\theta=\pi-x$ and $\theta=\pi+x$. In this range $2x$ in $\theta$ around $\theta=\pi$, chiral symmetry is restored. As chemical potential is increased further, this range increases and at some critical value of $\mu_{\rm Q}$, there is no such discontinuity for the pseudoscalar condensate and \ac{cp} symmetry is restored.

We then explored the impact of axions on the \ac{eos} for charge-neutral, beta-equilibrated quark matter to examine various characteristics of hybrid \ac{ns}s, which was our second primary motivation. Notably, a related study in Ref. \cite{Lopes:2022efy} investigated the stability of hybrid stars using the \ac{njl} model to describe quark matter with axions. Our work extends such studies in several ways. For the \ac{hqpt}, we modeled hadronic matter using a generalized Walecka model within quantum hadrodynamics, incorporating nonlinear meson interaction terms along with a quartic $\omega^2\zbf \rho^2$ term—specifically, the NL3${\omega\rho}$ model—including hyperons. On the quark matter side, we employed a three-flavor \ac{njl} model incorporating axions. Unlike Ref. \cite{Lopes:2022efy}, which used a Maxwell construction, we adopted a Gibbs construction for \ac{hqpt}. This approach allows for stable hybrid \ac{ns}s containing quark matter in a mixed phase with hyperonic matter, with or without axions, while satisfying modern astrophysical constraints from NICER X-ray observations and gravitational wave (GW) data from the GW170817 event. However, in the scenario where axion effects are maximal ($\theta = \pi$) and quark matter lacks vector repulsion ($G_v = 0$), the maximum mass constraint is not met. When $\theta=\pi$ and $G_v=0$, the star becomes unstable. Introducing a vector interaction in this case, however, enables the existence of stable hybrid NSs with a quark matter core, where the quark matter can exist either in a pure phase or in a mixed phase with hyperonic matter, surrounded by an outer core of hyperonic matter—all while satisfying the astrophysical maximum mass constraints.

The other important novel aspect of the present investigation lies in the study of \ac{nro}s of such hybrid \ac{ns}s with an axionic quark matter core. We focused our attention to quadruple fundamental modes here. We have performed the analysis of $f$ mode oscillations within the Cowling approximation which, in general, is not a good approximation for \ac{ns}s which are less compact. However, it turns out that such an approximation is not too bad for  more compact \ac{ns}s as is the case considered here for  stars with a quark matter core. It is observed that while $f$ modes get enhanced due to presence of hyperons as well as quark matter; the enhancement is particularly large in the presence of axions as they lead to a larger quark core with possibility of quark matter both in a pure phase or in a mixed phase with the hyperonic matter. The presence of quark matter as well as hyperons soften \ac{eos} which results in a substantial enhancement of the $f$ mode frequency as compared to the canonical nucleonic \ac{ns}s. Thus, a detection of the enhanced $f$ mode frequency could be indicative of non-neucleonic degrees of freedom in neutron star matter. 

The present study paves the way to further investigation in many different directions. Regarding the \ac{qcd} phase structure, it will be very interesting to include the effects of color-superconducting phase in the presence of axions. For two flavor quark matter, in fact, this has been recently attempted \cite{Murgana:2024djt, Balkin:2020dsr}. In the presence of axions, novel superconducting phases can arise in the pseudoscalar channel also making a rich phase structure for dense quark matter. In a three flavor scenario, the color superconductivity for the charge neutral matter with axions along with related gapless phases will be phenomenologically quite interesting particularly in the context of cooling of \ac{ns}s through axion emission \cite{Gomez-Banon:2024oux}. It will also be worth investigating the effect of strong magnetic field on such hybrid \ac{ns}s. In the context of \ac{ns}'s \ac{nro} modes, it will be interesting to go beyond the Cowling approximation to include metric perturbations leading to the quasi-normal modes which can be, in general, complex giving rise to damping of different oscillation modes. We have focused our attention to the $f$ modes here. It will be very interesting to study the gravity modes as well. Apart from the cold \ac{ns}s, quadruple oscillations also occur in newly born \ac{ns}, after \ac{ns} merger which are system where the temperature effects are important. Inclusion of temperature, entropy distribution in \ac{ns}, neutrino trapping effects can further be explored. Building on our work here, incorporating other effects like rotation and superfluidity will help in understanding the \ac{nro}s from such hybrid \ac{ns}s more thoroughly. Such studies will also be of interest to the structure of newly born protoneutron stars, \ac{ns} mergers particularly for the later case to see whether and how it gives insights to \ac{qcd} critical point in \ac{qcd} phase diagram in such gravity assisted collisions. We leave these interesting problems to near future works.
%
%
\begin{acknowledgments}
DK would like to thank School of Physical Sciences, NISER, Bhubaneswar for a visit where this work was initiated. He would also like to express his gratitude for the warm hospitality extended to him at Kamala Nibas, Bhubaneswar where part of the present work was done.
\end{acknowledgments}

\bibliography{zz_cpv}
\end{document}